\documentclass[11pt]{JHEP3}
\usepackage{amsmath,amssymb}
\usepackage{graphicx}

\newcommand{\al}{\alpha}

\newcommand\Tr{{\rm Tr}}

\def\Re{\mathrm{Re}\, }
\newcommand{\atopfrac}[2]{\genfrac{}{}{0pt}{}{#1}{#2}}
\newcommand{\f}[2]{\frac{#1}{#2}}
\newcommand{\K}[1]{{\rm \bf K}\mbox{$\left({#1}\right)$}}
\newcommand{\E}[1]{{\rm \bf E}\mbox{$\left({#1}\right)$}}

\newcommand{\pint}{\makebox[0pt][l]{\hspace{3.4pt}$-$}\int}
\newcommand{\ko}[1]{\left( #1 \right)}
\newcommand{\kko}[1]{\left[ #1 \right]}
\newcommand{\kkko}[1]{\left\{ #1 \right\}}
\newcommand{\ket}[1]{\left| #1 \right\rangle}

\newcommand{\bmt}[1]{{{\mbox{\boldmath$ #1 $}}}}
\newcommand{\komoji}[1]{\mbox{$#1$}}

\def\oneloop{{\rm 1\mbox{\scriptsize -}loop}}
\def\A{{\mathcal A}}
\def\B{{\mathcal B}}
\def\tlambda{{\widetilde \lambda}}

\def\AdS{{AdS${}_5$}}
\def\S{{S${}^5$}}

\def\AdSS{{AdS${}_5\times {}$S${}^5$}}
\def\AdSs{{AdS${}_4\times {}$S${}^2$}}
\def\Nf{${\mathcal N}=4$}
\def\H{{\mathcal H}}
\def\Tr{{\rm Tr\,}}

\def\const{{\rm const.}}
\def\L{{\mathcal L}}
\def\O{{\mathcal{O}}}
\def\calS{{\mathcal{S}}}
\def\calE{{\mathcal{E}}}
\def\calJ{{\mathcal{J}}}

\def\ds{\displaystyle}
\def\N{{\mathcal N}}
\def\C{{\mathcal C}}
\def\A{{\mathcal A}}
\def\pa{\partial}
\def\eq{\equiv}
\def\maru{\mbox{\scriptsize $\bigcirc$}}

\title{Open Spinning Strings and AdS/dCFT Duality}
\author{Keisuke Okamura\\
Department of Physics, 
Faculty of Science, 
University of Tokyo, 
Bunkyo-ku, 
Tokyo 113-0033, 
Japan\\
E-mail: \email{okamura@hep-th.phys.s.u-tokyo.ac.jp}
}
\author{Yastoshi Takayama\\
Department of Physics, 
Graduate School of Science, 
Osaka University, 
Toyonaka, 
Osaka 560-0043, 
Japan\\
E-mail: \email{takayama@het.phys.sci.osaka-u.ac.jp}
}
\author{Kentaroh Yoshida\\
Theory Division, 
Institute of Particle and Nuclear Studies, 
High Energy Accelerator Research Organization (KEK), 
Tsukuba, 
Ibaraki 305-0801, 
Japan\\
E-mail: \email{kyoshida@post.kek.jp}
}

\abstract{We consider open spinning string solutions on an {\AdSs}-brane
(D5-brane) in the bulk {\AdSS} background. By taking account of the
breaking of $SO(6)_{\rm R}$ to $SO(3)_{\rm H}\times SO(3)_{\rm V}$ due
to the presence of the AdS-brane, the open rotating string ansatz is
discussed. We construct the elliptic folded/circular open string solutions 
in the $SU(2)$ and the $SL(2)$ sectors, 
so that they satisfy the appropriate boundary conditions.  
On the other hand, in the
$SU(2)$ sector of the gauge theory, we compute the matrix of anomalous
dimension of the defect operator, which turns out to be the
Hamiltonian of an open integrable spin chain. Then we consider the
coordinate Bethe ansatz with arbitrary
number of impurities, and compare the boundary condition of the Bethe
wavefunction with that of the corresponding open string solution.
We also discuss the Bethe ansatz for the open $SL(2)$ spin chain 
with several supports from the string theory side. 
Then, in both $SU(2)$ and
$SL(2)$ sectors, we analyze the Bethe equations in the thermodynamic
limit and formulate the `doubling trick' on the Riemann surface
associated with the gauge theory.  }


\preprint{hep-th/0511139\\
UT-05-16\\
OU-HET 546\\
KEK-TH-1030}

\begin{document}

\section{Introduction}

Now we have a laboratory to check the AdS/CFT correspondence
\cite{M} in a non-BPS region by considering a large spin limit
\cite{BMN,GKP2,FT}. In this story, the classical rotating string solutions
with energies analytic in the 't Hooft coupling divided by the large-spin 
\cite{GKP2,FT,FT2,ART, Arutyunov} (for a review, see
\cite{review}) correspond to composite operators with definite scaling
dimensions in a super Yang-Mills (SYM) theory. The
diagonalization of the matrix of anomalous dimension can be performed by
using the Bethe ansatz because it is remarkably represented by an
integrable spin chain Hamiltonian \cite{MZ}.

This type of AdS/CFT duality in the non-BPS regime is 
investigated in each of subsectors of the full superconformal group
$PSU(2,2|4)$, which is closed under the operator
mixing. The $SU(2)$ \cite{BMSZ,BFST,KMMZ}, 
$SL(2)$ \cite{BFST,BS,SL2,KZ,ST-coh}, 
$SU(2|3)$ \cite{su(2|3),Stefanski:2005tr} sectors 
as well as $PSU(2,2|4)$ \cite{BS} are completely closed under the operator
mixing. The $SO(6)$ \cite{MZ,BMSZ} and 
$SU(3)$ sector \cite{ST-coh,HL,SU(3)} has been 
shown to be closed under the one-loop renormalization. 
In particular, the $SU(2)$ sector has been well studied.  
The higher-loop contributions are discussed in \cite{SS,BDS}, and the full-loop 
Bethe ansatz are also proposed in \cite{BDS}.
In both the BMN (near-BPS) and the Frolov-Tseytlin (far-from-BPS) sectors, 
the results of the perturbative computation in the SYM side have seen a remarkable agreement with
those in the string side up to the two-loop level, and the disagreement appears from the three-loop level. 

On the other hand, the classical integrability of type IIB superstring on the  {\AdSS} \cite{BPR} plays a key role in the correspondence  \cite{DNW}. The matching of the spectra and the integrable structures between the string  sigma models and the spin chains are confirmed up to and including the two-loop effects in some cases \cite{AS} (See also \cite{Arutyunov:2005nk}).
It is possible to capture the corrections of the half-integer powers of the 't Hooft coupling by investigating the integrable supercoset model $Osp(2m+2|2m)$, which gives the classical string equation of motion in the $SU(2)$ sector in the classical limit \cite{MP}. 

It is an interesting problem to generalize the correspondence to the open string cases, 
and indeed the concern with this topic has been growing.
In order to consider the open string sector, we need to insert some D-branes or orientifolds in the bulk {\AdSS}\,. 
One of such examples is the D3-D7-O7 system, for which the AdS/CFT duality in the near-BPS region is studied in 
\cite{BGMNN,Imamura}, and 
in the far-from BPS region in \cite{Stefanski,CWW1,CWW2, EM}. 
Another example is the open strings on giant gravitons, for which the duality in the near-BPS 
region is studied in \cite{BHLN,TT} and in the far-from BPS region in \cite{BV,BCV}. 
In addition, a set of long-range Bethe ansatz equations for open quantum strings on {\AdSS} are presented and indeed 
diagonalized in bosonic $SU(2)$ and $SL(2)$ sectors in the near pp-wave limit in \cite{MS}.  

Except the cases mentioned above, the {\AdSs} case is 
interesting to study. In this paper, we consider open strings 
on the {\AdSs}-brane (D5-brane) in the {\AdSs}\,. 
This setup is proposed by Karch
and Randall \cite{KR}. This AdS-brane leads to a three-dimensional
defect in four-dimensional Minkowski spacetime. The 3-5 strings supply
defect fields which couple to {\Nf} SYM fields. The presence of
the AdS-brane breaks the isometry (4D conformal ${}\times SO(6)_{\rm R}$
symmetry), $SO(2,4)\times SO(6) \to SO(2,3) \times SO(3)_{\rm H}\times
SO(3)_{\rm V}$\,, where the $SO(3)_{\rm H}$ denotes the isometry of the
S$^2$ part of the {\AdSs} and the $SO(2,3)$ implies the three-dimensional
conformal group\,.  The theory on the defect is superconformal and the
action has been constructed by DeWolfe-Freedman-Ooguri \cite{DFO}. The
superconformality in the non-abelian case has been shown in
\cite{EGK}. The BMN operator correspondence for open strings in this
setup has been shown by Lee and Park \cite{LP}. Then the anomalous
dimension matrix for the defect composite operators consisting of
$SO(6)$ scalars has been computed by DeWolfe and Mann \cite{DM}, and it
is represented by an integrable open $SO(6)$ spin chain Hamiltonian. In
our previous works we concentrate on the $SO(3)_{\rm H}$ sector by
truncating the $SO(6)$ sector.  In \cite{STY1}, the matching at
the level of the effective action is shown by using a coherent state method 
\cite{Kruczenski} (For a review of the effective action approach, 
see \cite{review-coh}). In \cite{STY2}, by focusing upon the open pulsating string 
in the $SO(3)_{\rm H}$ sector, the agreement between the energy and the anomalous 
dimension has been shown at the two-loop level.  
It has also been shown that the discrepancy starts from the three-loop level.
In the gauge theory side, the `doubling trick' has been discussed
on the Riemann surface. The doubling trick also works well for 
the energy of the open pulsating string solution.\\

We continue to consider the large-spin limit of the AdS/dCFT
correspondence that has been discussed
in \cite{STY1,STY2}. In this paper we discuss
rotating string solutions in the holomorphic sectors such as the $SU(2)$ and
the $SL(2)$ sectors, rather than non-holomorphic sectors such as the $SO(6)$ and
the $SO(3)_{\rm H}$ sectors. Recall that the $SO(6)_{\rm R}$ symmetry is spontaneously 
broken to $SO(3)_{\rm H}\times SO(3)_{\rm V}$\,. Then the
four-dimensional conformal symmetry $SO(2,4)$ is also broken to the 
three-dimensional conformal symmetry $SO(2,3)$\,.

Firstly, we propose the rotating string ansatz for the open string case
and specify the boundary conditions to be satisfied. 
As such open string solutions, we found an elliptic folded and an elliptic 
circular solutions in the $SU(2)$ sector, including a rational circular solution 
as a limiting case of the elliptic circular one.  In the $SL(2)$ sector, we found 
an elliptic folded solution.  
We also discuss the generic formulae for the one-loop string energy and the ratio 
of two spins, both in the $SU(2)$ and the $SL(2)$ 
sectors.  

Secondly, we discuss the SYM dual for
these solutions. In the $SU(2)$ sector, we compute the
matrix of anomalous dimension of the corresponding defect operators. 
It is shown to have the same structure as the Hamiltonian of an open $SU(2)$ integrable 
spin chain.  We use the coordinate Bethe ansatz, and discuss the 
Bethe wavefunction on the open spin chain.  The boundary condition of the resulting Bethe wavefunction 
turns out to be the Dirichlet one, that is fairly consistent with the boundary condition for 
its string dual whose endpoints sit on the D5-brane.  

We also comment on the $SU(3)$ sector. In contrast to the closed string case, 
in both the string and the gauge theory sides, 
the $SU(3)$ sector is not closed even at the one-loop level in our setup,

\vspace*{0.5cm}
This paper is organized as follows: In section 2 we discuss a rotating
string ansatz for open strings in the $SU(2)$ and the $SL(2)$ sectors and find 
solutions satisfying the boundary conditions supplied by the D5-brane.   
 In section 3 we compute the matrix of
anomalous dimension of the defect composite operator in the $SU(2)$ sector. 
The result is represented by an open spin chain
Hamiltonian with integrable boundaries. 
In section 4 we perform the Bethe ansatz via the coordinate approach, obtaining the 
Bethe wavefunction and the dispersion relation with arbitrary number of impurities.  
Then we discuss the boundary condition for the wavefunction, and compare with that 
for the open string solution in the string theory side.
In section 5 we consider the elliptic solutions for the open spin chain, which can be obtained 
from the symmetric two-cut solutions of the closed spin chain case via the `doubling trick' technique.
This technique is formulated and presented to give the halved version of the `double contour' and 
the `imaginary root' solution, which are well explored in the closed chain case.  These solutions for the 
closed cases are `opened' to give the elliptic solutions of the open spin chain associated with our dCFT setup.  
We then solve the Bethe equation in the scaling limit, obtaining a generic 
formulae for the one-loop anomalous dimension and the filling fraction, as well as the 
endpoints of the Bethe strings.    %
Section 6 is devoted to a conclusion and discussions. 
In Appendix \ref{app:3-loop formula} we solve the Riemann-Hilbert problem in the elliptic sector, and 
present a generic formula for the anomalous dimension at the three-loop level.  Some basic facts of 
complete elliptic integrals and the integral formulae are also summarized.

\section{Open Spinning Strings}

We are interested in open rotating string solutions. 
To find them we consider the holomorphic sector, rather than 
non-holomorphic sector where pulsating string or rotating pulsating
string solutions are allowed. In the open string case, 
the presence of D-branes leads to a spontaneous symmetry breaking, and so 
we need to take account of it into the rotating string ansatz. 
As a matter of course, boundary conditions should be also satisfied. 

First we will fix the coordinate system that is available in our setup.  
Then rotating string ansatze are proposed both in $SU(2)$ and $SL(2)$ sectors to 
find open rotating string solutions with two spins satisfying the boundary conditions.  
In the $SU(2)$ sector, we will find the elliptic folded, the elliptic circular and the rational circular open string 
solutions.  In the $SL(2)$ sector, on the other hand, the elliptic folded string solution is allowed.

\subsection{The Coordinate System}

We consider the near-horizon limit of the D3/D5 system.  The AdS/CFT study for this system has been developed in \cite{KR,DFO}.  The probe D5-brane lives on an {\AdSs} subspace of {\AdSS}, and the spacetime coordinates occupied by these branes are as follows:
\begin{center}
\begin{tabular}{c|cccccccccc}
{} &$x_{0}$ &$x_{1}$ &$x_{2}$ &$x_{3}$ &$x_{4}$ &$x_{5}$ &$x_{6}$ &$x_{7}$ &$x_{8}$ &$x_{9}$ \\
\hline
D3 &\maru &\maru &\maru &$\times$ &$\times$ &$\times$ &\maru &$\times$ &$\times$ &$\times$  \\
D5  &\maru &\maru &\maru &\maru &\maru &\maru &$\times$ &$\times$ &$\times$ &$\times$ \\
\end{tabular}
\end{center}
Then the near-horizon geometry is described by the coordinates $(t_{\rm M},\vec{y})\eq (x_{0},x_{1},x_{2})$, $x\eq x_{6}$, $u$ (radial coordinate) and the five angles $(\Psi,\phi, \varphi, \chi,\varsigma)$:
\begin{alignat}{5}
x_{3}&=u\cos\Psi\,\sin\phi\,\cos\varphi\,,&\qquad x_{4}&=u\cos\Psi\,\sin\phi\,\sin\varphi\,,&\qquad  x_{5}&=u\cos\Psi\,\cos\phi\,,\cr
x_{7}&=u\sin\Psi\,\sin\chi\,\cos\varsigma\,,&\qquad x_{8}&=u\sin\Psi\,\sin\chi\,\sin\varsigma\,,&\qquad  x_{9}&=u\sin\Psi\,\cos\chi\,.
\end{alignat}
In this setup the D5-brane sits at
$x=\Psi =0$\,, filling the AdS${}_4$ defined by the coordinates
$u,~t_{\rm M},~\vec{y}$ and wrapping the S${}^2$ parameterized by
$\phi$ and $\varphi$\,.  The metric for the near-horizon geometry in this coordinate system is 
\begin{align}
ds^2_{\rm AdS_5} &= R^2\left[
\frac{du^2}{u^2}+u^2(-dt_{\rm M}^2 + d\vec{y\,}^2 +dx^2)\right] \,,
\qquad \vec{y}=(y^1,y^2)\,, \label{metricAdS}\\ 
ds^2_{\rm S^5} &= R^2\Bigl[d\Psi^2+\cos^2\Psi\left(d\phi^2 
+ \sin^2\phi\,d\varphi^2\right) + \sin^2\Psi\left(d\chi^2 
+ \sin^2\chi\,d\varsigma^2\right) \Bigr]\,, \label{metricS}
\end{align}
with $R^4 = 4\pi\al'{}^2g_{\rm s}N$\,. \\

We should remind that the physical target-space interpretation of the solutions
depends on a particular choice of the coordinates.  As is the case with the closed string, for our purpose to study the
open strings on the {\AdSs}, the following global coordinate system are useful :  
\begin{alignat}{4}
Y_1  + i Y_2&\eq s_{1}\, e^{i\phi_{1}}= \sinh \rho \, \cos \theta \, e ^{i \phi _1 }\,,\qquad & X_1  + i X_2&\eq r_{1}\, e^{i\varphi_{1}}= \sin \gamma \, \cos \psi \, e ^{i \varphi _1 }\, ,\\
Y_3  + i Y_4&\eq s_{2}\, e^{i\phi_{2}}= \sinh \rho \, \sin \theta \, e ^{i \phi _2 }\, , \qquad &X_3  + i X_4&\eq r_{2}\, e^{i\varphi_{2}}= \sin \gamma \, \sin \psi \, e^{i \varphi _2 }\,  ,\\
Y_5  + i Y_0&\eq s_{0}\, e^{it}= \cosh \rho \, e ^{i t}\,.\qquad &X_5  + i X_6&\eq r_{3}\, e^{i\varphi_{3}}= \cos \gamma \, e ^{i \varphi _3 } \,.
\end{alignat}
Here $Y_{P}$ $\ko{P=0,\dots,5}$ and $X_{M}$ $\ko{M=1,\dots,6}$ are the embedding coordinates of {\AdS} and {\S}, respectively .  
For these coordinate system, the metric of the {\AdSS} is written as
\begin{align}
  ds^2 _{\left( {{\rm AdS}_5 } \right)}  &=  R^{2}\kko{{d\rho ^2  - \cosh ^2 \rho \,dt^2  + \sinh ^2 \rho \left( {d\theta ^2  + \cos ^2 \theta \,d\phi _1 ^2  + \sin ^2 \theta \,d\phi _2 ^2 } \right)}} \,, \label{metric-AdS}\\ 
  ds^2 _{\left( {\rm S^5 } \right)}  &=R^{2}\kko{{d\gamma ^2  + \cos ^2 \gamma \,d\varphi _3 ^2  + \sin ^2 \gamma \left( {d\psi ^2  + \cos ^2 \psi \,d\varphi _1 ^2  + \sin ^2 \psi \,d\varphi _2 ^2 } \right)}} \label{metric-S}
\end{align}
for the {\AdS} and the {\S} part respectively.  

Now let us see where the D5-brane is sitting in the present coordinate system (\ref{metric-AdS}) and (\ref{metric-S}).
Note that the metric (\ref{metricAdS}) can be obtained from (\ref{metric-AdS}) by the following transformation,
\begin{eqnarray}
&& Y_1 = i u\,t_{\rm M}\,, \qquad 
Y_i = u\,x^i 
\qquad \mbox{where} \quad x^i=(\vec{y},x)~~(i=2, 3, 4)\,,  \\ 
&& Y_0 = \frac{i}{2u}\Bigl[1-u^2(1-\vec{x}^2+t_{\rm M}^2)\Bigr]\,, \quad 
Y_5  = \frac{1}{2u}\Bigl[1+u^2(1+\vec{x}^2-t^2_{\rm M})\Bigr]\,,
\end{eqnarray}
 which is embedded in the six-dimensional flat space,
\begin{eqnarray}
ds^2 = R^2\left(-dY_0^2 - dY^2_5 + \sum_{i=1}^{4}dY^2_i\right) 
\quad \mbox{with} \quad Y_0^2+Y_5^2-\sum_{i=1}^4Y_i^2 = 1\,.
\label{metric3} 
\end{eqnarray}
Thus in the AdS space, we may take the position of the D5-brane as $\phi_2 =0$ without loss
of generality.  In the {\S} part, we may identify the location of the brane as $\psi=\varphi_{3}=0$.

Finally, the metric of embedding space of the {\AdSs} where our D5-brane (in the near-horizon limit) lives is given by
\begin{equation}
ds^{2}_{\left( {{\rm AdS}_4 }\times {\rm S}^{2} \right)}
=R^{2}\kko{{d\rho ^2  - \cosh ^2 \rho \,dt^2  + \sinh ^2 \rho \left( {d\theta ^2  + \cos ^2 \theta \,d\phi _1 ^2} \right)}+{d\gamma ^2  + \sin ^2 \gamma \,d\varphi _1 ^2}}\,.
\end{equation}

\subsection{String Sigma Model Action of Open String on ${{\rm AdS}_{4}\times {\rm S}^{2}}$} 

The string sigma model for an open string is given by 
\begin{equation}
S = \frac{\sqrt{\lambda}}{2}
\int\!d\tau\int^{\pi}_0\!\frac{d\sigma}{\pi}\bigl[
\L_{\rm AdS_5} + \L_{\rm S^5}\bigr]\,,  \label{action} 
\end{equation}
where $R^2/\al'$ is rewritten in terms of the 't Hooft coupling $\lambda$
via the fundamental relation of the AdS/CFT correspondence,
$R^4/\al'{}^2 = Ng^2 \equiv \lambda$, and the Lagrangians for the {\AdS} and the {\S} are given by
\begin{align}
\L_{\rm AdS_5} &= -\frac{1}{2}\eta^{PQ}\partial_a Y_P \partial^a Y_Q 
+ \frac{1}{2}\widetilde{\Lambda}\left(\eta^{PQ}Y_PY_Q + 1\right)\,, \\ 
\L_{\rm S^5} &= -\frac{1}{2}\partial_a X_M \partial^a X_M 
+ \frac{1}{2}\Lambda\left(X_MX_M-1\right)\,.
\end{align}
 The index $a$ denotes
the world-sheet coordinates $(\tau,\sigma)$ and $\eta^{PQ}=(-1,+1,+1,+1,+1,-1)$.
The functions $\widetilde{\Lambda}$
and $\Lambda$ are the Lagrange multipliers which depend on $\tau$ and
$\sigma$\,. As in the closed string case, we use the usual conformal
gauge and so the action (\ref{action}) is supplemented
with the Virasoro conditions to ensure the vanishing of the total
two-dimensional energy-momentum tensor:
\begin{align}
&\eta^{PQ}\left(\dot{Y}_P\dot{Y}_Q + Y_P'Y_Q'\right) + \dot{X}_M\dot{X}_M 
+ X'_MX'_M = 0\,, \\ 
&\eta^{PQ}\dot{Y}_P {Y}_Q'+ \dot{X}_M X'_M=0\,,
\end{align}
where the dot and the prime denote $\pa_{\tau}$ and $\pa_{\sigma}$, respectively.  

We can obtain the sigma model charges as
\begin{eqnarray}
S_{PQ} = \frac{\sqrt{\lambda}}{2}\int^{\pi}_0\frac{d\sigma}{\pi}\,\left(
Y_P\dot{Y}_Q - Y_Q\dot{Y}_P\right)\,, \quad 
J_{MN} =
\frac{\sqrt{\lambda}}{2}\int^{\pi}_0\frac{d\sigma}{\pi}\,\left(
X_M\dot{X}_N - X_N\dot{X}_M\right)\,. \label{cartan}
\end{eqnarray}
The only difference from the closed string case is the range of $\sigma$, which now runs from $0$ to $\pi$.  Then the AdS energy $E$, the AdS spins $S_1$ and $S_2$\,, and the S$^5$ spins $J_1$, $J_2$ and $J_{3}$ associated with the shifts of $t$\,, $\phi_1$\,, $\phi_2$\,, $\varphi_1$ $\varphi_2$ and
$\varphi_3$\,, respectively, are given by 
\begin{alignat}{4}
E &= S_{50} = \sqrt{\lambda}\,\mathcal{E}\,, \quad &
S_1 &= S_{12} = \sqrt{\lambda}\,\mathcal{S}_1\,, \quad &
S_2 &= S_{34} = \sqrt{\lambda}\,\mathcal{S}_2\,; \\ 
J_1 &= J_{12} = \sqrt{\lambda}\,\mathcal{J}_1\,, \quad &
J_2 &= J_{34} = \sqrt{\lambda}\,\mathcal{J}_2\,, \quad &
J_3 &= J_{56} = \sqrt{\lambda}\,\mathcal{J}_3\,,
\end{alignat}

Now we consider the boundary effects.  The action (\ref{action}) has an $SO(2,4)\times SO(6)$ global symmetry, but 
it is spontaneously broken to 
$SO(2,3) \times SO(3)_{\rm H} \times SO(3)_{\rm V}$ due to the presence of the AdS-brane.  
We remark that the charge densities of $S_2$, $J_{2}$ and $J_3$ should vanish at the
boundaries. Here we should also comment on the three-dimensional conformal group $SO(2,3)$. 
The four-dimensional conformal group $SO(2,4)$ is generated by 
\begin{eqnarray}
S_{\mu\nu} = M_{\mu\nu}\,, \quad S_{\mu 4} = \frac{1}{2}(K_{\mu}-
 P_{\mu})\,, \quad S_{\mu 5}=\frac{1}{2}(K_{\mu} + P_{\mu})\,, \quad 
S_{54} = D\,,
\end{eqnarray}
where $\mu,~\nu=0,1,2,3$\,. The generators
$M_{\mu\nu}\,,~P_{\mu}\,,~K_{\mu}\,,~D$ are representation of 4D rotation, 4D translation,
special conformal and dilatation respectively.  One can identify the standard spin
with $S_1 = S_{12} = M_{12}$\,, and the second (conformal) spin with
$S_2 = S_{34} = \frac{1}{2}(K_3-P_3)$\,.  The conformal energy is the
rotation generator in the 0-5 plane, i.e., the AdS energy
$E=S_{05}=\frac{1}{2}(K_0+P_0)$\,.  As we have mentioned above, 
the four-dimensional conformal group
(dim($SO(2,4)$)=15) is spontaneously broken to the three-dimensional
conformal group (dim($SO(2,3)$)=10). Concretely speaking, $S_{34}$ and
$S_{35}$ as well as $M_{03}\,,~M_{13}\,,~M_{23}$ are broken. In
particular, the broken $S_{34}$ and $S_{35}$ imply the broken $P_3$ and
$K_3$.  As a result, $S_2$ is also broken.  These results reflects the
vanishing $\phi_2$ on the AdS-brane.  \\

The crucially different point is, as compared to the closed string case, that the open strings on a D-brane (without coupling to a gauge field on the brane) should satisfy either Neumann
or Dirichlet boundary condition at their endpoints. Since the D-brane sits on $\phi_2 =0$ in the AdS${}_{5}$ and $\psi =\varphi_{3}=0$ in the {\S} as we have already mentioned, the angular variables $\phi_2$, $\psi$ and $\varphi_{3}$ should vanish at both $\sigma
=0$ and $\pi$ :
\begin{eqnarray}
\left.\phi_2(\tau,\sigma)\vphantom{\f{}{}}\right|_{\sigma=0,\pi} = 0 \qquad \mbox{and} 
\qquad \left.\psi(\tau,\sigma)\vphantom{\f{}{}}\right|_{\sigma=0,\pi}=
\left.\varphi_{3}(\tau,\sigma)\vphantom{\f{}{}}\right|_{\sigma=0,\pi} = 0\,. \label{bcy}
\end{eqnarray}
We are interested in classical solutions of the sigma model 
(\ref{action}) satisfying (\ref{bcy})\,. 

Here recall that an open string is reduced to half of a closed string
via the doubling trick. If we find open string solutions, then those
may be related to the solutions of the closed string sigma model via the
doubling trick. But the inverse argument does not hold generally. All of the closed
string solutions do not lead to open string solutions. Some of the closed
string solutions can be `opened' to satisfy the boundary conditions
(\ref{bcy}) and others not.  We argue that if and only if the folding number 
of the folded string or the winding number of the circular string takes the 
even integer value, those closed string solutions can be split into two open 
string solutions in our brane setup.  
This condition for the doubling trick to work can also be found in the 
gauge theory side.  We will make a sufficient argument on this point later in section \ref{sec:DT-gauge}.

\subsection{Rotating String Ansatz and Sigma Model Charges}

We are interested in the string solutions whose energy can be expanded regularly in powers of $\lambda/J^{2}$. They can be obtained by imposing appropriate ansatz on the string sigma model action.

\subsubsection*{Field Equations and Virasoro Constraints}

In the elliptic sectors where the angular variables $\phi_{a}$ and $\varphi_{i}$ depend only on $\tau$, the original {\AdSS} Neumann-Rosochatius system reduces to the sum of two Neumann systems, namely one for the {\AdS}  and the other for the {\S} \cite{review}.  Then the equations of motion in terms of the global coordinates completely decouple into two Neumann systems; one for the $(\rho,\theta)$-system,
\begin{alignat}{2}
&\mbox{for}\quad \rho\,,&\quad &\rho'' -  \sinh \rho\,\cosh \rho\,\ko{\kappa^2 + \theta '{}^2 - \omega_1^2 \cos^2 \theta - \omega_2^2 \sin^2 \theta }   =0\,,\label{EOM1}\\
&\mbox{for}\quad \theta\,,&\quad &\ko{\sinh^2 \rho\,\theta'}'   -  \ko{\omega_1^2 - \omega_2^2} \sinh^2 \rho\,\sin \theta \,\cos \theta  =0\,, 
\end{alignat}
and the other for the $(\gamma,\psi)$-system,
\begin{alignat}{2}
&\mbox{for}\quad \gamma\,,&\quad &\gamma'' -  \sin \gamma\,\cos \gamma\,\ko{w_3^2 + \psi'{}^2 - w_1^2 \cos^2 \psi - w_2^2 \sin^2 \psi }=0\,,\\
&\mbox{for}\quad \psi\,,&\quad &\ko{\sin^2 \gamma\,\psi'}'   - \ko{w_1^2 - w_2^2} \sin^2 \gamma\,\sin \psi \,\cos \psi  =0\,.\label{EOM2}
\end{alignat}
Coupled as above at the level of the field equations, those four angular variables are subject to one nontrivial Virasoro constraint,
\begin{align}
&\rho'{}^2  - \kappa^2 \cosh^2 \rho   + \sinh^2\rho \ko{ \theta'{}^2 +\omega_1^2 \cos^2 \theta  + \omega_2^2\sin^2 \theta } \cr
&{}+\gamma'{}^2 +  w_3^2 \cos^2\gamma  + \sin^2\gamma\ko{\psi'{}^2 +w_1^2 \cos^2\psi + w_2^2 \sin^2\psi} = 0 \ . \label{Virasoro}
\end{align}

In both $SU(2)$ and $SL(2)$ sectors, there is a rotating string solution called a folded string, which has a topology of an interval.  The folded string in the $SU(2)$ sector and the one in the $SL(2)$ sector are related via an `analytic continuation' of global coordinates, as we will see below.  
In the $SU(2)$ sector, there is another kind of elliptic solution known as a circular string which has a topology of a circle, and it generally winds around the great circle of the {\S}.  This circular solution includes a rational circular solution as a certain limit.  

In the following, we will see each of the solutions in turn.

\subsubsection*{Elliptic Folded/Circular Open Strings in ${SU(2)}$ Sector}

The elliptic folded open string in the $SU(2)$ sector is stretched in the $\psi$-direction within the equator $\gamma=\pi/2$ of {\S} and rotates around its collocated endpoints with angular momentum (or `spin') $J_{2}$.  The endpoints move along another orthogonal great circle of {\S} with the spin $J_{1}$.  

On the other hand, the elliptic circular open string wraps around the equator $\gamma=\pi/2$ of {\S}.  In this circular case, two endpoints have the $\psi$-coordinate in common at $\psi=0$, which is the location of the D5-brane in {\S}, whereas the $\varphi_{i}$-coordinate may be different between them, thus it can be distinguished from the elliptic circular closed string.


\FIGURE{
\centerline{
\includegraphics[width=0.85\textwidth]{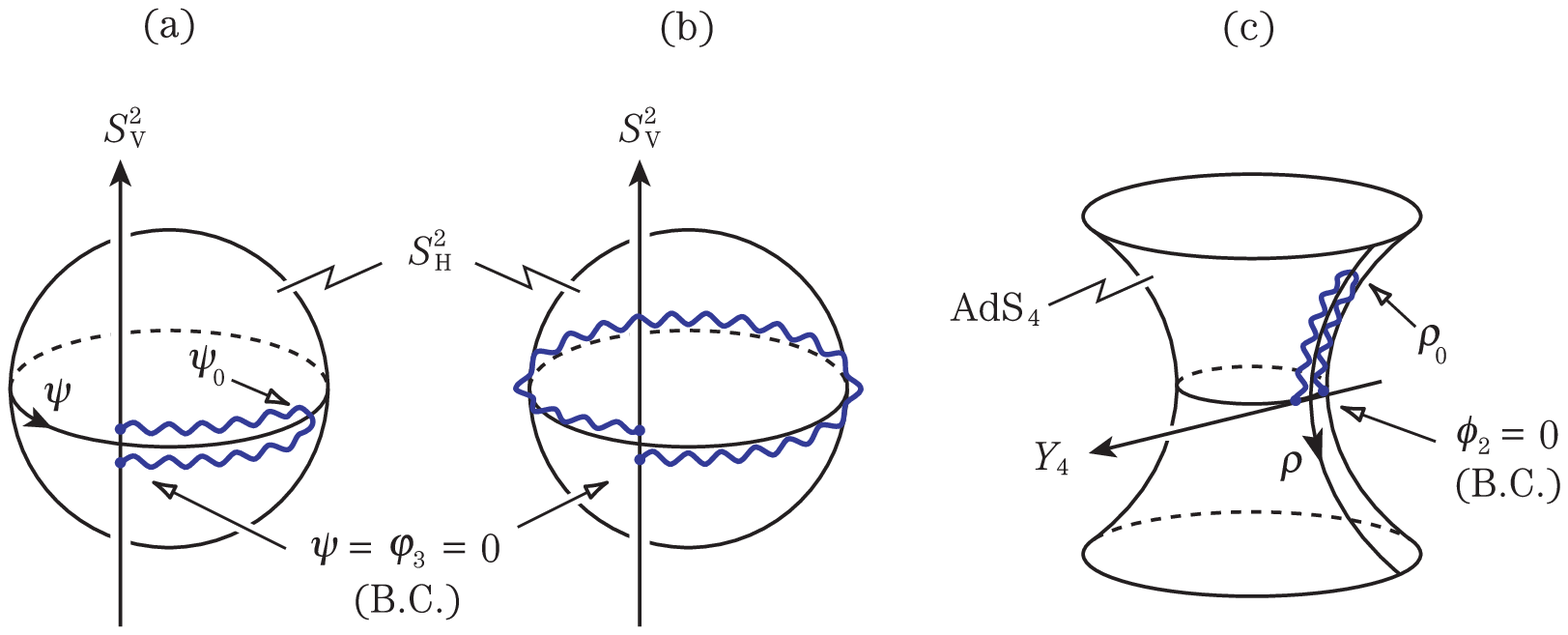}
\caption{\small Open spinning string solutions in far-from-BPS (Frolov-Tseytlin) sector; (a): elliptic folded open string in the $SU(2)$ sector, (b): elliptic circular open string in the $SU(2)$ sector, (c): elliptic folded open string in the $SL(2)$ sector.  Here ``B.C.'' stands for the boundary condition.
}
\label{fig:open-spinning-string}
}}

The rotating string ansatz for both types of the elliptic strings (folded/circular) in the $SU(2)$ sector is as follows:
\begin{alignat}{6}\label{ansatz-SU(2)}
\mbox{{\AdS} side :}&\quad &
{}&\rho=0\,,\quad &
{}&\theta=0\,,\quad &
{}&\phi_1=0\,,\quad &
{}&\phi_2=0\,,\quad &
{}&t=\kappa\tau\,;\\
\mbox{{\S} side :}&\quad &
{}&\gamma=\f{\pi}{2}\,,\quad &
{}&\psi=\psi\ko{\sigma}\,,\quad &
{}&\varphi_1=w_1\tau\,,\quad &
{}&\varphi_2=w_2\tau\,,\quad &
{}&\varphi_{3}=0\,.
\end{alignat}
This ansatz is endowed with the following boundary condition,
\begin{equation}\label{BC-SU(2)}
\psi\ko{0}=\psi\ko{\pi}=0\,,
\end{equation}
instead of the periodic condition of the closed case.  With the ansatz (\ref{ansatz-SU(2)}), the {\AdS}- and {\S}-charges become
\begin{alignat}{3}
\calS_1  &=0 \,,
 &\qquad \calS_2  &=0\,,
 & \qquad \calE &= \kappa\,;\\
\calJ_1  &=w_1\int_0^{\pi} \f{d\sigma}{\pi}\,\cos^2\psi \,,
 &\qquad \calJ_2  &=w_2\int_0^{\pi} \f{d\sigma}{\pi}\,\sin^2\psi\,,
 & \qquad \calJ_3 &= 0\,.
\end{alignat}
Integrating the field equation (\ref{EOM2}) once, we obtain
\begin{equation}\label{sin}
\psi'{}^{2}=\ko{w_{2}^{2}-w_{1}^{2}}\ko{C-\sin^{2}\psi}\,,
\end{equation}
where $C$ is an integration constant.  The arguments we have held so far are valid for both folded and circular string solutions; the topology of the open string depends completely on the value of $C$.

Let us first see the $0<C<1$ case.  In this case we can set $C\eq \sin^{2}\psi_{0}$. Then from (\ref{sin}), we obtain
\begin{equation}
\psi'=\pm\sqrt{\ko{w_{2}^{2}-w_{1}^{2}}\ko{\sin^{2}\psi_{0}-\sin^{2}\psi}}\,,\label{psi'}
\end{equation}
where we assumed $|w_{2}|\geq |w_{1}|$.  
This is an elliptic folded open string folded in the interval $\psi\in [0,\psi_{0}]$.
The charges are calculated by using the condition (\ref{psi'}) as
\begin{equation}
\left\{
\begin{array}{ccc}
\calJ_{1}\\
\calJ_{2}\\
\end{array} 
\right\}
=\left\{
\begin{array}{ccc}
w_{1}\\
w_{2}\\
\end{array} 
\right\}\int_{0}^{\psi_{0}}\f{2n\, d\sigma}{\pi}
{{\left\{
\begin{array}{ccc}
\cos^{2}\psi (\sigma)\\
\sin^{2}\psi (\sigma)\\
\end{array} 
\right\}
} \mathord{\left/ {\vphantom {{\left\{
\begin{array}{ccc}
\cos^{2}\psi (\sigma)\\
\sin^{2}\psi (\sigma)\\
\end{array} 
\right\}
} {\sqrt{\ko{w_{2}^{2}-w_{1}^{2}}\ko{\sin^{2}\psi_{0}-\sin^{2}\psi}}
}}} \right. \kern-\nulldelimiterspace} {\sqrt{\ko{w_{2}^{2}-w_{1}^{2}}\ko{\sin^{2}\psi_{0}-\sin^{2}\psi (\sigma)}}}}\,.
\end{equation}
which can be expressed via complete elliptic integrals of the first and the second, as we will see in section \ref{sec:energy-and-spins}.  Here the integer $n$ counts how many times the open string repeats the period $\psi=0\to \psi_{0}\to 0$.  For example, the folding number 1, 3/2, 2 corresponds to those have the shape of ``$>$'', ``Z'', ``$\Sigma $'', respectively.  
Hence the $n$-folded open string consists of $2n$ segments.  For example, when $n=1$, 
$\psi(\sigma)$ increases from 0 to $\psi_{0}$ when $\sigma$ increases 
from 0 to $\pi/2$, and then decreases back toward $\psi=0$
when $\sigma$ goes on to $\pi$.  This is just the half trajectory of the closed string case.  
The radial coordinates for an elliptic folded open string in the $SU(2)$ sector are given by
\begin{align}
r_1\ko{\sigma ; n ; x}&=\cos\psi\ko{\sigma}={\mathrm{dn}}\mbox{$\ko{\f{2n}{\pi}\, \K{1-x}\,\sigma,1-x}$}\, ,\label{radii-F-1}\\
r_2\ko{\sigma ; n ; x}&=\sin\psi\ko{\sigma}=\sqrt{1-x}\, {\mathrm {sn}}\mbox{$\ko{\f{2n}{\pi}\,\K{1-x}\,\sigma,1-x}$}\, ,\label{radii-F-2}
\end{align}
where $\sigma$ runs from $0$ to $\pi$.
Here $x\in [0,1]$ is the elliptic moduli for the folded string which has the following geometrical interpretation,%
\footnote{\, Our definition of $x$ is related to the moduli $q$ used in \cite{review} by the relation $x=1-q$.}%
\begin{equation}
x=\cos^{2}\psi_{0}\,,
\end{equation}
where $\psi_{0}$ is the maximum value of $\psi$ for the folded string in {\S}.  \\

On the other hand, an open elliptic circular string appears when $C>1$.  In this case, the radii are given by
\begin{align}
r_1\ko{\sigma ; n ; y}&=\cos\psi\ko{\sigma}={\mathrm{cn}}\mbox{$\ko{\f{2n}{\pi}\, \K{y}\,\sigma,y}$}\label{r1-c}\, ,\\
r_2\ko{\sigma ; n ; y}&=\sin\psi\ko{\sigma}={\mathrm {sn}}\mbox{$\ko{\f{2n}{\pi}\,\K{y}\,\sigma,y}$}\label{r2-c}\, ,
\end{align}
with $y\in [0,1]$.  This time the elliptic moduli $y$ controls the distribution of the energy density in the open string, and the integer $n$ counts how many times the open string winds around the equator of $\gamma=0$ in the $\psi$-direction.  As is obvious from (\ref{r1-c}) and (\ref{r2-c}), only even-$n$ cases can satisfy the B.C. (\ref{BC-SU(2)}) for our open strings.  We also note that the position vector $\vec{r}=(r_{1},r_{2})$ for both the folded and the circular strings are invariant under the shift $\sigma\mapsto \sigma +\pi$ for the even-$n$ cases, which we can see from Eqs.\,(\ref{periodicity-sn})-(\ref{periodicity-dn}).  \\%

Clearly, in both folded and circular cases, the doubling trick in terms of the energy-spin relation reads
\begin{equation}\label{DT-string}
\delta_{\rm (o)}(J_{1}^{\rm (o)},J_{2}^{\rm (o)})=\f{1}{2}\delta_{\rm (c)}(J_{1}^{\rm (c)},J_{2}^{\rm (c)})\,,
\qquad J_{i}^{\rm (c)}=2J_{i}^{\rm (o)}\quad \ko{i=1,2}\,,
\end{equation}
where the subscripts ``c'' and ``o'' indicate the charges of closed and open strings, respectively.

\subsubsection*{Rational Circular Open String in ${SU(2)}$ Sector}

A rational circular string appears in the limit $y\to 0$ of the elliptic circular string:
\begin{equation}
\ko{r_1\ko{\sigma ; n ; y},\, r_2\ko{\sigma ; n; y}}\quad \xrightarrow[y\to 0]{}\quad \ko{\cos n\sigma,\, \sin n\sigma}\,.
\end{equation}
Obviously, only for even $n$, this solution makes sense as an open string satisfying the boundary condition (\ref{BC-SU(2)}).  This type of string solution can be also realized as the equal two-spin limit of the so-called `constant radii' ($r_{i}=\const$) solution in the $SU(2)$ sector via a global $SU(2)$ rotation, see the arguments below (\ref{string-rational}).

\subsubsection*{Elliptic Folded Open String in ${SL(2)}$ Sector}

In the $SL(2)$ sector, we consider the following two-spin ansatz:
\begin{alignat}{6}
\mbox{{\AdS} side :}&\qquad &
{}&\rho=\rho\ko{\sigma}\,,\qquad &
{}&\theta=0\,,\qquad &
{}&\phi_1=\omega_1\tau\,,\qquad &
{}&\phi_2=0\,,\qquad &
{}&t=\kappa\tau\,;\\
\mbox{{\S} side :}&\qquad &
{}&\gamma=\f{\pi}{2}\,,\qquad &
{}&\psi=0\,,\qquad &
{}&\varphi_1=w_1\tau\,,\qquad &
{}&\varphi_2=0\,,\qquad &
{}&\varphi_3=0\,.\label{ansatz-SL(2)}
\end{alignat}
With this ansatz, we can construct the elliptic folded open string which is stretched in the $\rho$-direction of {\AdS} and rotates around its collocated endpoints with spin $S_{1}$. The endpoints also move in the {\S} side with the spin $J_{1}$, along one of the great circle.  In contrast to the $SU(2)$ case, the boundary conditions supplied by the D-brane are automatically satisfied within the ansatz (\ref{ansatz-SL(2)}).  
The charges in this sector are given by
\begin{alignat}{3}
\calS_1  &=\omega_1\int_0^{\pi} \f{d\sigma}{\pi}\,\sinh^2\rho~~\ko{\equiv\calS}\,,
 &\qquad \calS_2  &=0\,,
 & \qquad \calE &= \kappa\int_0^{\pi} \f{d\sigma}{\pi}\,\cosh^2\rho\,;\label{SL(2)-charges1}\\
\calJ_1  &=w_{1}~~\ko{\equiv\calJ}\,,
 &\qquad \calJ_2  &=0\,,
  & \qquad \calJ_3 &= 0\,.\label{SL(2)-charges2}
\end{alignat}
These charges can be written in terms of the moduli $x\in [1,\infty)$, which indicates the maximum value of $\rho$ (which we denote $\rho_{0}$) by the relation
\begin{equation}
x=\cosh^{2}\rho_{0}\,.
\end{equation}

One important observation \cite{ART,BFST} is that, when we reduce the equations of motion and the Virasoro constraint according to our ansatze for the $SU(2)$ (\ref{ansatz-SU(2)}) and the $SL(2)$ (\ref{ansatz-SL(2)}) sector, they map to themselves via the following transformations,
\begin{alignat}{7}\label{AC-angle}
\mbox{$SL(2)$ :}\qquad   &\rho\,\quad &  &\omega_{1}\,\quad &  &\kappa\,\quad &  &w_{1}\,&  \cr
{}  &\updownarrow & &\updownarrow &  &\updownarrow & & \updownarrow &\quad .\\
\mbox{$SU(2)$ :}\qquad  & i \psi\, \quad &  -&w_{2}\,\quad &  -&w_{1}\,\quad &   &\hspace{-0.1cm}{-\kappa}\, &\nonumber
\end{alignat}
This can be regarded as a kind of `analytic continuation' between the two metrics (\ref{metric-AdS}) and (\ref{metric-S}).  We can see that the charges in the $SL(2)$ sector are related to those in the $SU(2)$ sigma model via the following analytic continuation:  %
\begin{alignat}{5}\label{AC-charge}
\mbox{$SL(2)$ :}\qquad   &\calS_{1}\,\quad &  &\calE\,\quad &  &\calJ_{1}\,&  \cr
{}  &\updownarrow & &\updownarrow && \updownarrow &\quad .\\
\mbox{$SU(2)$ :}\qquad  & \calJ_{2}\, \quad & -&\calJ_{1}\, \quad & &\hspace{-0.1cm}{-\calE}\, &\nonumber
\end{alignat}
Note that, in this $SL(2)$ sector, the $\rho$-coordinates do not have to start with or end with $\rho=0$ as $\sigma$ increases from $0$ to $\pi$, but we impose these constraints, so that the open string approaches point-like (BPS) string in the limit $\calS-\calE\to 0$.

\subsection{One-loop String Energy and Ratio of Spins\label{sec:energy-and-spins}}

In the string theory side we expand the classical energy in powers of $\lambda$ divided by the large-spin squared, e.g., 
\begin{equation}
E\ko{J_{1},J_{2}}=J\kko{1+\f{\lambda}{J^{2}}\,\epsilon^{(1)}\komoji{\ko{\f{J_{2}}{J}}}+\f{\lambda^{2}}{J^{4}}\,\epsilon^{(2)}\komoji{\ko{\f{J_{2}}{J}}}+\dots}\,,
\end{equation}
with $J\eq J_{1}+J_{2}$ for the $SU(2)$ sector, and
\begin{equation}
E\ko{S,J}=S+J\kko{1+\f{\lambda}{J^{2}}\,\epsilon^{(1)}\komoji{\ko{\f{S}{J}}}+\f{\lambda^{2}}{J^{4}}\,\epsilon^{(2)}\komoji{\ko{\f{S}{J}}}+\dots}\,,
\end{equation}
with $S\eq S_{1}$ and $J\eq J_{1}$ for the $SL(2)$ sector.  

From now on we present the general expression for the one-loop energy $\delta_{\rm s}\eq \epsilon^{(1)}\lambda/J$, and then reduce it to obtain a particular string solution such as folded and circular string.  

\subsubsection*{ Generic Form}

The `ratio-of-spins' and the one-loop (in $\lambda$) string energy of the generic elliptic open string can be written as
\begin{equation}
\alpha_{\rm s}\ko{x}=\f{\E{x}+i \ko{\E{1-x}-\K{1-x}}}{\K{x}-i \K{1-x}}\,,\label{gen_as}
\end{equation}
and 
\begin{align}
\delta_{\rm s}\ko{n;x}&=-\f{n^{2}\lambda}{2\pi^{2}J}\ko{-\K{x}+i \K{1-x}}\cr
&\hspace{0.8cm}{}\times \kko{\E{x}+i \ko{\E{1-x}-\K{1-x}}\ko{-\K{x}+i \K{1-x}}}\,,\label{gen_ENE}
\end{align}
respectively, where the moduli $x$ of the complete elliptic integrals takes a real value in $[0,1]$.  Here the ratio-of-spins $\alpha_{\rm s}$ represents the ratio of $\calJ_{2}$ to the total spin $\calJ_{1}+\calJ_{2}$ for the $(J_{1},J_{2})$-solution in the $SU(2)$ sector, whereas in the $SL(2)$ sector where we consider $(S,J)$-solution, it plays the role of the ratio of $\calJ$ to $\calS$ multiplied by $-1$.

\subsubsection*{ Elliptic Folded Open String}

Let us see the case of the folded open string solution in the $SU(2)$ sector.
The ratio-of-spins and the one-loop (in $\lambda$) energy of the elliptic folded open string are obtained from the generic formulae (\ref{gen_as}) and (\ref{gen_ENE}) by analytic continuing them past $x=0$, which lead to
\begin{equation}
\al_{\rm fold}\ko{x}=1-\f{\E{1-x}}{\K{1-x}}\label{aF}
\end{equation}
for the ratio-of-spins, and 
\begin{align}
\delta_{\rm fold}\ko{n;x}=-\f{n^{2}\lambda}{2\pi^{2}J}\K{1-x}\kko{-\E{1-x}+x\K{1-x}}\,,\label{dF}
\end{align}
for the energy, respectively.  When $n=2$, the open string repeats the period of $\psi=0\to \psi_{0}\to 0$ twice, and the energy equals to that of the folded closed string solution whose $\psi$-coordinate changes $0\to \psi_{0}\to 0\to -\psi_{0}\to 0$ as $\sigma$ runs from $0$ to $2\pi$.  Namely, $\delta_{\rm fold}\ko{n=2;1-x_{0}}$ and $\al_{\rm fold}\ko{1-x_{0}}$ are equal to Eq.\,(2.7) of \cite{BFST}.  \\

The folded open string in the $SL(2)$ sector has almost the same expressions as (\ref{aF}) and (\ref{dF}), only to multiply an extra minus sign in both the energy and the ratio-of-spins, i.e.,
\begin{align}
\widetilde\alpha_{\rm fold}\ko{z}&=-1+\f{\E{1-z}}{\K{1-z}}\,,\label{aF2}\\
\widetilde\delta_{\rm fold}\ko{n;z}&=\f{n^{2}\lambda}{2\pi^{2}J}\K{1-z}\kko{-\E{1-z}+z\K{1-z}}\,.\label{dF2}
\end{align}
Again, $\widetilde\delta_{\rm fold}\ko{n=2;1-x_{0}}$ and $\widetilde\alpha_{\rm fold}\ko{1-x_{0}}$ are equal to Eq.\,(B.18) and (B.19) of \cite{BFST}.

\subsubsection*{ Elliptic Circular Open String}

The ratio-of-spins and the one-loop energy of the elliptic circular string are obtained by the following procedure.  First, let us analytically continue the generic function (\ref{gen_as}) and (\ref{gen_ENE}) past $x=1$, and perform a modular transformation making use of the relations (\ref{mod.trans-1})-(\ref{mod.trans-2}).  This gives
\begin{equation}
\al_{\rm circ}\ko{y}=\f{\E{y}-\ko{1-y}\K{y}}{y\K{y}}\label{ROS-ellip.circ}
\end{equation}
for the ratio-of-spins, and
\begin{equation}
\delta_{\rm circ}\ko{n;y}=\f{2n^{2}\lambda}{\pi^{2}J}\E{y}\K{y}\label{ENE-ellip.circ}
\end{equation}
for the one-loop energy, where the new moduli is related to the old one via $y=1/x\in [0,1]$.  

In this circular case, contrast to the folded case, the integer $n$ counts how many times the open string winds around the equator.   %
Thus $\delta_{\rm circ}\ko{n;y}$ is equal to the energy of the closed string solution with winding number $n$, which is $n^{2}$ times Eq.\,(D.9) of \cite{BFST}.\footnote{\, The ratio-of-spins $\alpha$ in Eq.\,(D.9) of \cite{BFST} is related to (\ref{ROS-ellip.circ}) via $\alpha=1-\al_{\rm circ}$.} 

\subsubsection*{ Rational Circular Open String}

Let us take the rational limit $y\to 0$ in the ratio-of-spins (\ref{ROS-ellip.circ}) and the one-loop energy (\ref{ENE-ellip.circ}).  We have
\begin{equation}\label{string-rational}
\al_{\rm circ}\ko{y}\quad \xrightarrow[y\to 0]{}\quad \f{1}{2}\,,\qquad 
\delta_{\rm circ}\ko{n,y}\quad \xrightarrow[y\to 0]{}\quad \f{n^{2}\lambda}{2J}\,,
\end{equation}
which is known as the `simplest circular string' solution in the $SU(2)$ sector.  This solution can be also realized as the $\al_{\rm s}\to 1/2$ limit of the `constant-radii string' solution, which follows from the following ansatz,
\begin{align}\label{constant-radii}
X_{1}+iX_{2}=a_{1}\, e^{i\ko{w_{1}\tau+n_{1}\sigma}}\,,\qquad 
X_{3}+iX_{4}=a_{2}\, e^{i\ko{w_{2}\tau+n_{2}\sigma}}
\end{align}
with constants $a_{i}$.  
The constant-radii solution can be regarded as the rational limit of the elliptic circular solution when 
\begin{equation}\label{const-radii}
n_{1}=-n_{2}\eq n\,,\qquad \mbox{and}\qquad w_{1}=w_{2}\,.
\end{equation}
In this case, the constant-radii-string (\ref{constant-radii}) has the same one-loop energy and the ratio-of-spins as (\ref{string-rational}), see Fig.\,\ref{fig:del-vs-ros} in the next subsection.  In fact, they are related through a global $SU(2)$ rotation.  We can say that the constant-radii solution can be `opened' to satisfy the boundary conditions (\ref{BC-SU(2)}) only when (\ref{const-radii}) are satisfied with an even winding number $n$; for unequal two-spin cases, the doubling trick does not work well in general.

\subsection{Spectroscopy of One-loop Energy via Ratio of Spins}


\FIGURE{
\centerline{
\includegraphics[width=0.9\textwidth]{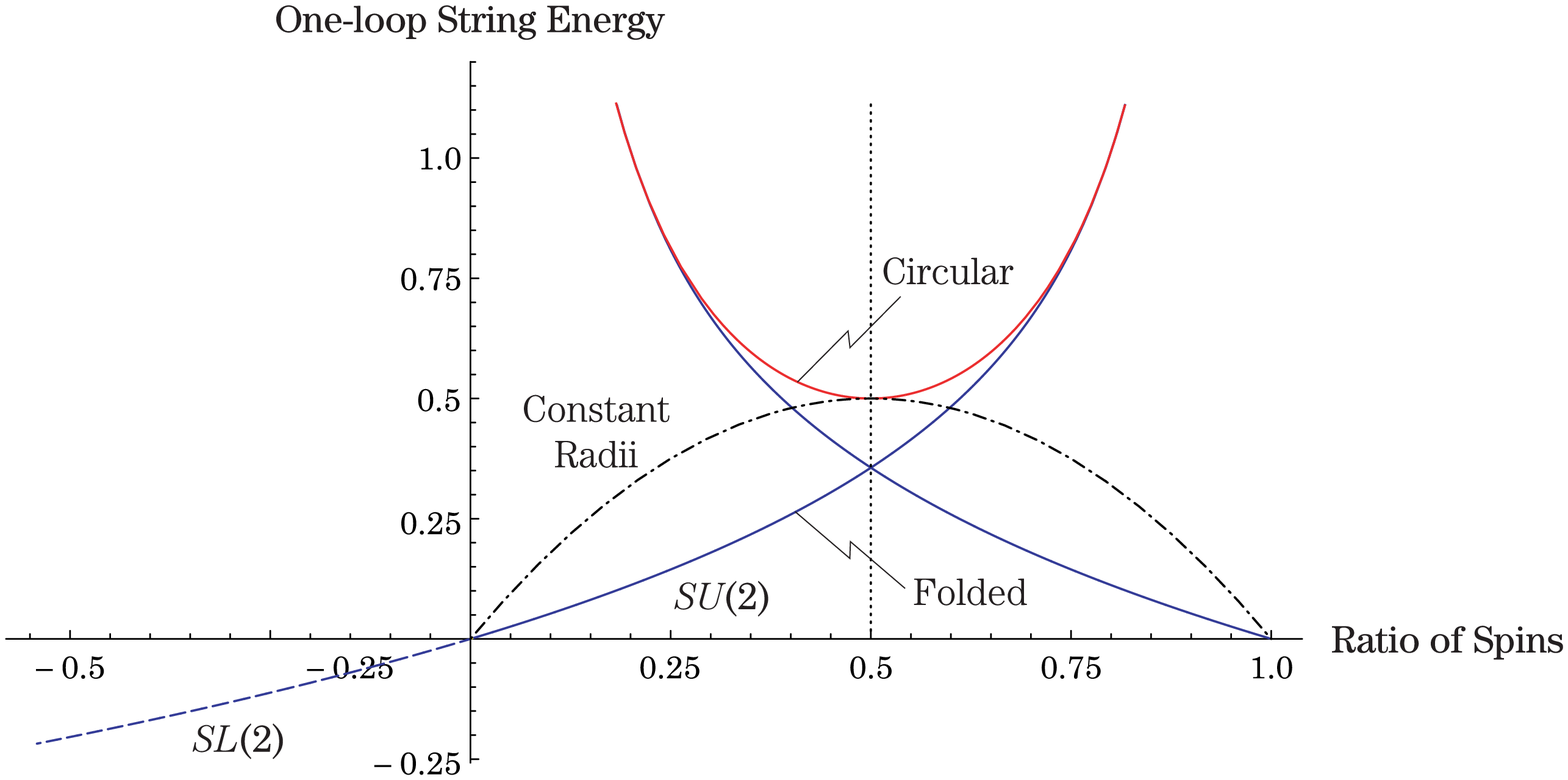}
\caption{\small One-loop string energy vs. ratio of spins.  The one-loop energy $\delta_{\rm s}$ (divided by $\lambda/L$) for each string solution is shown as a function of $\alpha_{\rm s}$.  Here the folding number for the elliptic folded open string is $n=2$, the winding number for the circular open string is $n=2$ and the winding number for the rational circular open string is $n=2$. To obtain an open string solution from a closed one via the doubling trick, the folding/winding number of the closed string has to be even.
}
\label{fig:del-vs-ros}
}}


\FIGURE{
\centerline{
\includegraphics[width=0.95\textwidth]{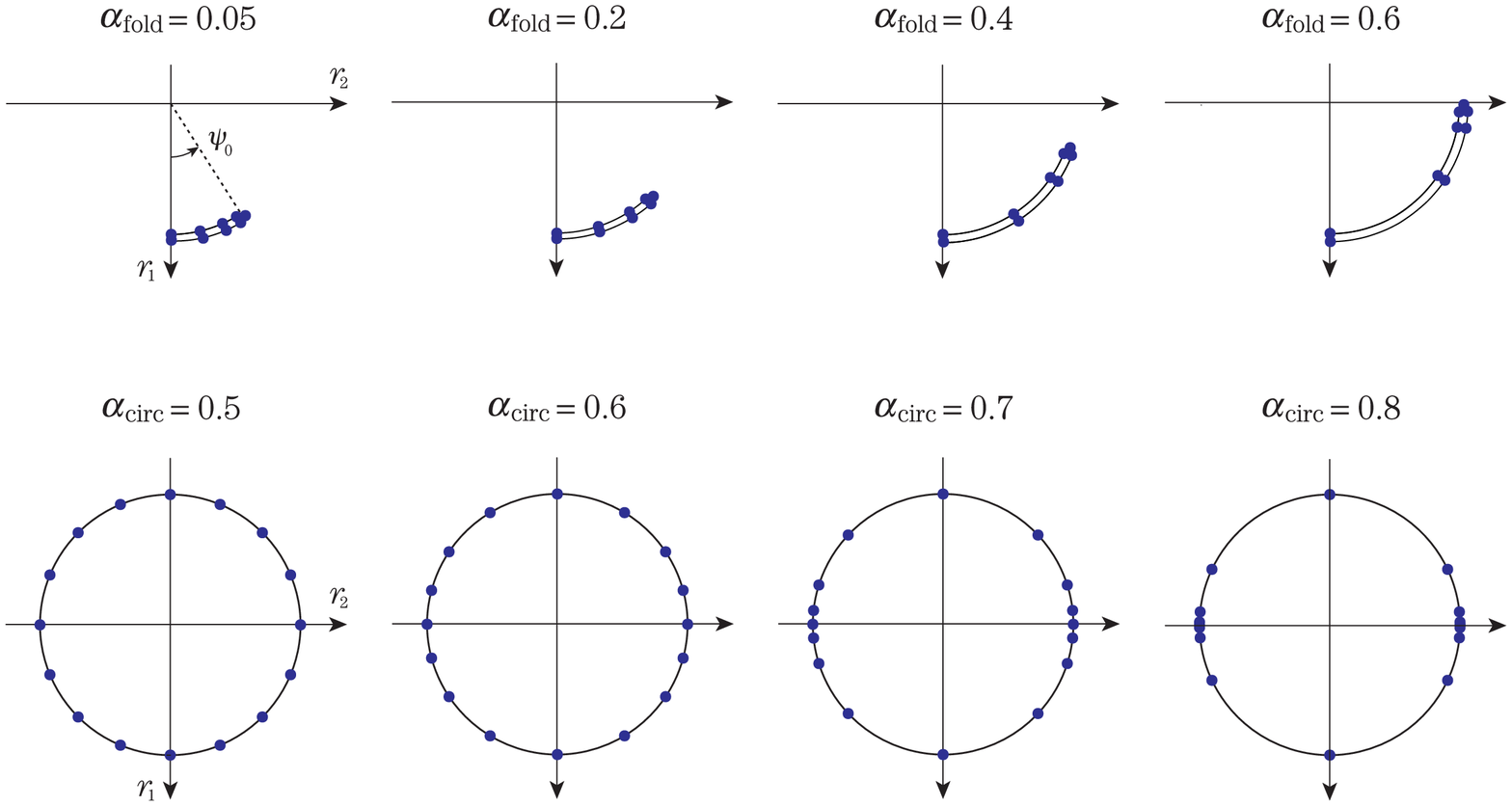}
\caption{\small Distribution of the energy density of an open folded (upper) and an open circular (lower) spinning string at various values of the ratio-of-spins.  The upper figures represent the folded open string with folding number $n=1$, where the small circles are plotted for $\sigma=k\pi/8$ $(k=0,\dots,8)$.  The lower ones represent the circular open string with winding number $n=1$, where the small circles are plotted for $\sigma=k\pi/16$ $(k=0,\dots,16)$.
}
\label{fig:density}
}}

Let us see how one-loop energy behaves as a function of the ratio-of-spins, for each solution in turn.\footnote{\, We thank K. Sakai for useful discussions on this subject.}  See Figs.\,\ref{fig:del-vs-ros} and \ref{fig:density}.  
When $\alpha_{\rm s}$ is zero, ($x= 1$ in terms of the elliptic moduli,) the folded open string shrinks to a point-like string on the D5-brane where the string energy does not depend on $\lambda$, i.e., BPS.  As $\alpha_{\rm s}$ increases from zero, the point-like open string begins to stretch into the $\psi$-direction.  Starting with the almost point-like (BMN) string near $x=1$, let us follow the curve of $\delta_{\rm fold}$ as a function of $\alpha_{\rm fold}$ (`Folded' branch) in the direction of increasing $\alpha_{\rm fold}$.  The ratio of $J_{2}$ to $J_{1}$ increases as we follow up the curve.  Then the energy density (or tension) of the folded open string begins,  as $\alpha_{\rm fold}$ approaches one, to concentrate on around the turning point where $\psi$-coordinate takes the maximum value $\psi_{0}$.  This fact can be immediately seen from the $\sigma$-dependence of the radii $r_{i}$ (\ref{radii-F-1}) and (\ref{radii-F-2}), and it may well be considered as the effect of `centrifugal force' associated with the dominant $J_{2}$-rotation.  At the critical point $x=0$ where $\alpha_{\rm fold}=1$, the folded open string can make a transition into the shape of a circular open string if and only if the folding number is even, say $2n$.  

Next let us follow the `Circular' branch, which is another branch allowed in the physical region $x<0$ of $\delta_{\rm s}\ko{2n;x}$, in the direction of lowering $\alpha_{\rm circ}$.  Then the energy density begins to uniformly distribute all over the open string.  In the limit $x\to -\infty$ where $\alpha_{\rm circ}=1/2$, the density takes the same value at any $\sigma\in [0,\pi]$.  In this case, the string rotates with equal two spins \---- this is the rational (simplest) circular open string.  

Finally let us follow the `Folded' branch where the ratio-of-spins takes a negative value.  This branch corresponds to $x\in [1,\infty)$, and the energy vs. ratio-of-spins curve in this region represents the elliptic folded open string in the $SL(2)$ sector.  %

\section{Open Integrable Spin Chain from dCFT : Perturbative Calculation\label{sec:one-loop-SU(2)}}

In this section we discuss the anomalous dimension matrix for the $SU(2)$  sector in the defect conformal field theory (dCFT).
First we briefly review the dCFT.
Then we calculate the one-loop anomalous dimension matrix for the $SU(2)$ sector, which turns out to be an open $SU(2)$ spin chain Hamiltonian with diagonal boundaries.

\bigskip
Let us make a short review on the dCFT \cite{DFO}.
The total action of the dCFT is given by the action of $\mathcal{N}=4$ SYM,
\begin{align}
S_{\mathcal{N}=4} &= \frac{1}{g^2} \int d^4x \,{\rm Tr}\, 
\Bigg\{\frac{1}{2} F_{\mu\nu} F^{\mu\nu} 
- i \overline\lambda^\alpha \gamma^\mu D_\mu \lambda^\alpha 
\cr
&\hspace{3.5cm}{}+  D_\mu X^i D^\mu X^j
- \frac{1}{2} [X^a, X^b]^2 
+ C_{\alpha\beta}^i \overline\lambda^\alpha [X^i, \lambda^\beta] \Bigg\} \,,
\end{align}
together with the actions for the defect fundamental fields $(q, \Psi)$:
\begin{align}
S_3 &= S_{\rm kin} + S_{\rm yuk} + S_{\rm pot}\,, \label{S3}\\
S_{\rm kin} &= \frac{1}{g^2}\int\!\!d^3x\Bigl[
(D^k q^m)^{\dagger}D_k q^m -i\overline{\Psi}^a\rho^kD_k\Psi^a
\Bigr]
\,, \\
S_{\rm yuk} &= \frac{1}{g^2}\int\!\!d^3x\Bigl[
i\overline{\Psi}^aP_+\lambda_{am}q^m -i\overline{q}^m\overline{\lambda}_{ma}P_+\Psi^a 
+ \overline{\Psi}^a\sigma^A_{ab}X_{\rm V}^{A}\Psi^a
\Bigr]\,,  \\
S_{\rm pot} &= \frac{1}{g^2}\int\!\!d^3x\Bigl[
\overline{q}^m X_{\rm V}^AX_{\rm V}^Aq^m 
+ i\epsilon_{IJK}\overline{q}^m\sigma_{mn}^{I}X_{\rm H}^JX_{\rm H}^Kq^n
\Bigr]  \nonumber\\
&\quad {} + \frac{1}{g^2}\int\!\!d^3x\left[
\overline{q}^m\sigma^I_{mn}(D_3 X_{\rm H}^I)q^n 
+ \frac{1}{4}\delta(0)\, {\rm Tr}(\overline{q}^m\sigma^I_{mn}q^n)^2
\right]\,. \label{Spot}
\end{align}
Here the covariant derivative is defined as $D_\mu * = \partial_\mu * - i [A_\mu,*]$ for the ${\mathcal N}=4$ adjoint fields and $ D_k* = \partial_k* - i A_k*$ for the fundamental defect fields.
The coupling to the defect fields breaks the $SO(6)_{\rm R}$ symmetry to $SO(3)_{\rm H} \times SO(3)_{\rm V}$.
As a result, the {\Nf} vector multiplet is decomposed into the following two multiplets,
\begin{alignat}{3}
& \mbox{3D vector multiplet}&~:&\quad 
\{A_k, P_+\lambda^{\alpha}, X_{\rm V}^A, D_3X_{\rm H}^I\}\,, \\
& \mbox{3D hyper multiplet}&~:&\quad 
\{A_3, P_-\lambda^{\alpha}, X_{\rm H}^I, D_3 X_{\rm V}^A\}\,,
\end{alignat}
where the suffixes $k\, (=0,1,2)$ and $i\, (=1,\dots,6)$ of $SO(6)_{\rm R}$ are decomposed into ${\bf 3}$ of the $SO(3)_{\rm V}$ ($A=1,2,3$) and ${\bf 3}$ of the $SO(3)_{\rm H}$ ($I = 4,5,6$).  
The defect fields $q^m$ and $\Psi^a$ transform in $({\bf 2}, {\bf 1})$ and in $({\bf 1}, {\bf 2})$ of $SO(3)_{\rm H}\times SO(3)_{\rm V}$, respectively.  

\subsection{One-Loop Analysis of ${SU(2)}$ sector}

Now we concentrate on the $SU(2)$ sector. 
The $SU(2)$ defect operator consists of $J_{1}$  $Z$'s and $J_{2}$ $W$'s with two defect fields $q^{m}$ and $\overline{q}^{m}$ on both ends,
\begin{equation}
 \mathcal{O} = \overline{q}_1 Z^{J_1} W^{J_2} q_2 + \cdots,
\label{SU(2)op}
\end{equation}
where ``$\cdots$'' indicates the operator mixing.  
The `length' of the operator (\ref{SU(2)op}) is defined by its bare dimension, $L+1\eq J_1+J_2+1$, and the complex scalars are built out of $SO(6)$ scalars such that
\begin{equation}
Z\eq \ko{{X_{\rm H}^1+iX_{\rm H}^2}}/\sqrt{2}\,,\qquad 
W\eq \ko{{X_{\rm V}^3+iX_{\rm V}^4}}/\sqrt{2}\,.
\end{equation}
The chiral primary operator is given by
\begin{equation}
 \mathcal{O} = \overline{q}_1 Z^{L} q_2\,,
\label{CPO}
\end{equation}
which corresponds to the Bethe reference state when we diagonalize the anomalous dimension matrix by using Bethe ansatz techniques.  
It should be noted that, when we associate this system with an open spin chain system later, the definition of the `endpoints' of the open chain is \textit{not} the location of the first and the last ($L$-th) sites of the chain. Instead, we have to identify the endpoints with half-step past the first and the last sites (i.e., the `$1/2$-th' and the `$L+1/2$-th' sites) \cite{DM}.  

In the one-loop level analysis, the defect interactions give additional contributions to the anomalous dimension matrix, whereas the bulk interactions lead to the same result in the closed string case \cite{MZ}.
(See Figs.\,\ref{fig:diagram}  (a), (b) and (c))


\FIGURE{
\centerline{
\includegraphics[width=0.8\textwidth]{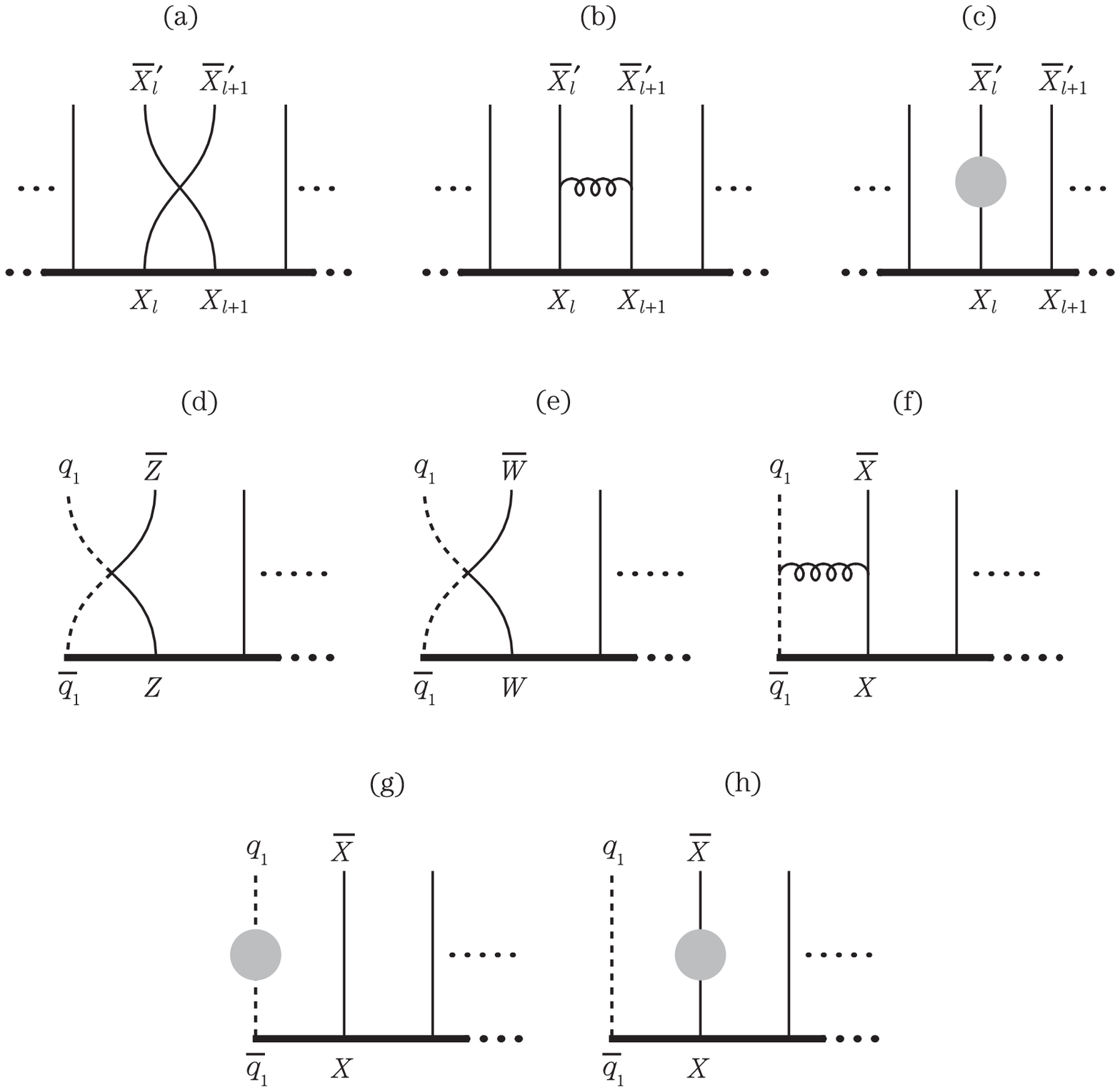}
\caption{\small The diagrams (a), (b) and (c) show the bulk interactions. The diagrams (d)-(h) contribute to the defect interactions.
}
\label{fig:diagram}
}}

Since $Z$ and $W$ have different flavor symmetries ($SO(3)_{\rm H}$ and $SO(3)_{\rm V}$ respectively), these fields give different contributions to the anomalous dimension through flavor-dependent defect interactions.
These flavor-dependent defect interactions (of our interest) come from (\ref{Spot}):
\begin{eqnarray}
 S_{\rm pot} = \frac{1}{g^2} \int d^3x
\Bigl[ \overline{q}^m \overline{W}W q^m
+ \overline{q}^m W \overline{W} q^m
+ \overline{q}^m \sigma^3_{mn} [\overline{Z},Z] q_n
+\cdots
\Bigr]. \label{fldep}
\end{eqnarray}

In order to obtain the additional contributions, let us concentrate on the left endpoint of (\ref{SU(2)op}) and consider the following correlation function
\begin{eqnarray}
 \Gamma_{\rm L}= \Big\langle \big[\overline{q}_1 X(0) \big]  \overline X(z_2)q_1(y_1) \Big\rangle. 
\end{eqnarray}
Here the adjoint scalar complex field $X$ denotes either $Z$ or $W$.
By considering the Callan-Symanzik equation, the one-loop anomalous dimension for $\Gamma_{\rm L}$ is obtained as
\begin{equation}
\gamma_{\rm L} \Gamma_{\rm L}^{\rm tree}=
 -\Lambda \frac{\partial}{\partial \Lambda}
\left[
\Gamma_{\partial}^{\rm \oneloop}
+\Gamma_{A}^{\rm \oneloop} + \Gamma_q^{\rm \oneloop}
\right]
 -\frac{\Lambda}{2} \frac{\partial}{\partial \Lambda} \Gamma_{X}^{\rm \oneloop}
 -\left[ \gamma_q + \frac{\gamma_X}{2} \right] \Gamma_{\rm L}^{\rm tree}.
\label{CSL}
\end{equation}
Each term in the above expression is evaluated as
\begin{alignat}{3}
\Lambda \frac{\partial}{\partial \Lambda} \Gamma_{\partial}^{\rm \oneloop}
&=&{}&\hspace{-1.7cm}{}
\left\{
\begin{array}{@{\,}cl@{\,}}
\ds \frac{\lambda}{8 \pi^2}  & \mbox{(four vertex for $q_1$ and $Z$ : Fig.\,1 (d))}\\
\vspace{-0.5cm}
&\cr
\ds  -\frac{\lambda}{8 \pi^2} & \mbox{(four vertex for $q_1$ and $W$ : Fig.\,1 (e))} \\
\end{array}
\right.\,,
\label{fldepCS}
\\
\Lambda \frac{\partial}{\partial \Lambda} \Gamma_{A}^{\rm \oneloop}
&=
\frac{\lambda}{16 \pi^2} &\qquad& \mbox{(gluon exchange : Fig.\,1 (f))}\,, \\
\Lambda \frac{\partial}{\partial \Lambda} \Gamma_q^{\rm \oneloop}
&=
-\frac{\lambda}{ 4\pi^2} &\qquad& \mbox{(self energy for $q^{m}$ : Fig.\,1 (g))}\,, \\
\Lambda \frac{\partial}{\partial \Lambda} \Gamma_{X}^{\rm \oneloop}
&=
\frac{\lambda}{8 \pi^2} &\qquad& \mbox{(self energy for $X$ : Fig.\,1 (h))}\,, \\
\gamma_q = \gamma_X
 &= 
 \frac{\lambda}{8 \pi^2} &\qquad& \mbox{(anomalous dimensions for $q^{m}$ and $X$)}, \label{q-X}
\end{alignat}
with $\lambda= g^2 N$.
(The common coordinate-dependent factor $\displaystyle \Gamma_{\rm L}^{\rm tree} = N \delta^3(y_1) \delta^4(z_{2})$ is ignored in these equations.)
The result (\ref{fldepCS}) comes from the flavor-dependent interactions (\ref{fldep}).
By inserting (\ref{fldepCS})-(\ref{q-X}) into (\ref{CSL}), the anomalous  dimension for $\Gamma_{\rm L}$ is obtained as
\begin{eqnarray}
 \gamma_{\rm L}
=
\left\{
\begin{array}{@{\,}cl@{\,}}
 \displaystyle 0 & \mbox{for } \, X=Z \\
 \displaystyle \frac{4 \lambda}{16\pi^2} & \mbox{for } \, X=W
\end{array}
\right.\,.
\end{eqnarray}
One also obtains the additional contribution to the anomalous dimension from the right endpoint of (\ref{SU(2)op}) in the same way.

\bigskip
\par
Hence the resulting anomalous dimension matrix for the $SU(2)$ sector is given by
\begin{align}
\Gamma_{SU(2)}
&=
\frac{\lambda}{16 \pi^2} \sum_{l=1}^{L-1} \H_{l,l+1}
+\frac{4\lambda}{16 \pi^2}\left(1- Q_1^W\right)
+\frac{4\lambda}{16 \pi^2}\left(1- Q_L^W\right),\label{GammaSU(2)}\\
&\mbox{with} \quad 
\H_{l,l+1}= 2I_{l,l+1} -2 P_{l,l+1}. \nonumber
\end{align}
The bulk part (the first term in (\ref{GammaSU(2)})) is the same as the closed $SU(2)$ spin chain case \cite{MZ} except the periodicity, and it is written in terms of the identity operator $I$ and the permutation operator $P$ as
\begin{equation}
 I_{j_{l}j_{l+1}}^{i_{l}i_{l+1}} 
= \delta^{i_l}_{j_l}\delta^{i_{l+1}}_{j_{l+1}}\,, \quad 
P_{j_{l}j_{l+1}}^{i_li_{l+1}} = \delta^{i_l}_{j_{l+1}}
\delta_{j_l}^{i_{l+1}}.\nonumber
\end{equation}
The projection operators $Q^W_{1,L}$ in the second and the third terms in (\ref{GammaSU(2)}) act on the first and the last ($L$-th) site of the open chain in the following way:
\begin{alignat}{3}
&Q_{1}^{W}\ket{W\cdots}=0\,,&\qquad &Q_{1}^{W}\ket{Z\cdots}=\ket{Z\cdots }\,,\label{projection-1}\\
&Q_{L}^{W}\ket{\cdots W}=0\,,&\qquad &Q_{L}^{W}\ket{\cdots Z}=\ket{\cdots Z}\,.\label{projection-L}
\end{alignat}
In the matrix notation, Eq.\,(\ref{GammaSU(2)}) can be rewritten as
\begin{align}
\Gamma_{SU(2)}
&=
\frac{\lambda}{16 \pi^2} \sum_{l=1}^{L-1} \H_{l,l+1}
+\frac{\lambda}{16 \pi^2} \left(4 \Sigma_1 + 4 \Sigma_L \right)\,,
\label{admSU(2)}
\\
&\mbox{with} \quad 
\H_{l,l+1}= I_{l,l+1} - \sum_{a=1}^3 \sigma^a_l \sigma^a_{l+1}\,,\nonumber
\end{align}
with
\begin{equation}\label{Sigma}
\Sigma_1 =
 \left[
 \begin{array}{@{\,}cc@{\,}}
  0 & 0 \\
  0 & 1
 \end{array}
 \right] (\otimes I_{2\times 2})^L ,
\qquad 
\Sigma_L =
 (I_{2\times 2} \otimes)^L 
\left[
 \begin{array}{@{\,}cc@{\,}}
  0 & 0 \\
  0 & 1
 \end{array}
 \right],
\qquad
I= {I_{2\times 2}}^{\otimes (L+1)}.
\end{equation}
The last two terms in (\ref{admSU(2)}) are the additional contributions coming from the defect interactions.
Thus the open spin chain Hamiltonian consists of the bulk part and the integrable boundaries, which ensure the integrability as will be shown below.  From the viewpoint of the integrability, the integrable boundaries satisfy the boundary Yang-Baxter relation in addition to the bulk Yang-Baxter relation.\footnote{\, This line of study has been done for the $\N=2$ $Sp(N/2)$ SYM theory in \cite{CWW2}}\\

The $\Gamma_{SU(2)}$ is shown to be an open $SU(2)$ integrable spin chain Hamiltonian with diagonal boundaries as follows.  
The most general Hamiltonian of an integrable 
$SU(n)$ open spin chain is given by \cite{Sklyanin,deVega:1993xi,Grisaru,Doikou:1998xi}
\begin{equation}
H_{SU(n)} =\sum_{l=1}^{L-1} \widetilde{\H}_{l,l+1}
+\frac{1}{4\xi_-}\frac{d}{d u}K^-_{1,(p)}(u,\xi_-)\bigg|_{u=0}
+\frac{{\rm tr}_0 \kko{K^+_{0,(p)}(0,\xi_{+}) \widetilde{\mathcal{H}}_{L,0}}}{{\rm tr}\, K^+_{(p)}(0,\xi_+)}\,,
\label{SU(n)open}
\end{equation}
together with  the $n \times n$ diagonal $K$-matrix
\begin{equation}
 K^{\pm}_{(p)} (u, \xi_{\pm}) =
  {\rm diag}(\underbrace{a^{\pm},\dots, a^{\pm}}_{p}, \underbrace{b^{\pm},\dots,b^{\pm}}_{n-p})\,, 
\end{equation}
with
\begin{alignat}{2}
 a^- &=i\xi_{-}+u\,,   &\qquad  b^- &=i\xi_{-}-u\,, \\
 a^+ &= i\ko{\xi_{+}-n}-u\,, &\qquad  b^+ &= i\xi_{+}+u
\end{alignat}
for any $p \in \{1,...,n-1\}$.  In the Hamiltonian (\ref{SU(n)open}), $K^+_{l,(p)}$ denotes the $K$-matrix which acts on the Hilbert space associated with the $l$-th site, where we assign $l=0$ to the auxiliary space.  The two-site (antiferromagnetic) Hamiltonian $\widetilde{\H}_{l,l+1}\eq \ko{P-I}_{l,l+1}/4$ acts on the product of the vector spaces $\mathbb C^{n}\times \mathbb C^{n}$.  
Here the arbitrary parameters $\xi_{\pm}$ may be regarded as
certain boundary magnetic fields.  They should be determined according to 
the setup of the dCFT and be evaluated with the perturbative calculation 
as we have just done for the $SU(2)$ sector.

The algebraic Bethe equations corresponding to the Hamiltonian (\ref{SU(n)open}) are given by 
\begin{eqnarray}
 1 &=& \bigg[e_{-2\xi_{-}+p}(u^{(p)}_i) e_{2\xi_{+}-p}(u^{(p)}_i) \delta_{p,k} +(1-\delta_{p,k}) \bigg] \nonumber\\
&&\times
	\prod_{j=1}^{M^{(k-1)}}
	 e_{-1}(u_i^{(k)} - u_j^{(k-1)})
	 e_{-1}(u_i^{(k)} + u_j^{(k-1)}) \nonumber\\
&&\times
	\prod_{j=1(j\neq i)}^{M^{(k)}}
	e_{2}(u_i^{(k)}-u_j^{(k)})
	e_{2}(u_i^{(k)}+u_j^{(k)}) \nonumber\\
&&\times
	\prod_{j=1}^{M^{(k+1)}}
	 e_{-1}(u_i^{(k)} - u_j^{(k+1)})
	 e_{-1}(u_i^{(k)} + u_j^{(k+1)}) \label{SU(n)BE}\\
&{\rm with}& \quad
e_m(u) \eq \frac{u+\frac{im}{2}}{u-\frac{im}{2}}, \nonumber
\end{eqnarray}
for $i=1,\dots, M^{(k)}$ and $k=1,\dots, n-1$.
Here we have defined $M^{(0)} = L$, $M^{(n)}=0$, and $u_i^{(0)}= u_i^{(n)}=0$.

Turning to the present dCFT case where $n=2$ and $p=1$, the Hamiltonian (\ref{SU(n)open}) translates to (up to a constant term),
\begin{align}\label{SU(2)-Hamiltonian}
H_{SU(2)}^{(p=1)}=-\f{1}{4}\kko{\sum_{l=1}^{L-1}\H_{l,l+1}+\f{2}{\xi_{-}}\Sigma_{1}+\f{2}{1-\xi_{+}}\Sigma_{L}}\,,
\end{align}
with $\Sigma_{1,L}$ defined by (\ref{Sigma}), and the Bethe equations (denoting $M\eq M^{(1)}$),
\begin{eqnarray}
 \left(\frac{u_i +i/2}{u_i -i/2}\right)^{2L}
=
\f{2u+i\ko{-2\xi_{-}+1}}{2u-i\ko{-2\xi_{-}+1}}\cdot \f{2u+i\ko{2\xi_{+}-1}}{2u-i\ko{2\xi_{+}-1}}
\prod_{j(\neq i)}^{M} \frac{u_i-u_j+i}{u_i-u_j-i}\cdot \frac{u_i+u_j+i}{u_i+u_j-i}\,,
\label{SU(2)BE}
\end{eqnarray}
for $i=1,\dots,M$.  The r.h.s.\ of (\ref{SU(2)BE}) consists of two boundary $S$-matrices defined at both ends of the chain, and the product of the bulk two-body $S$-matrices.  
It should be remarked that the boundary magnetic parameters $\xi_{\pm}$ cannot be determined just by assuming the integrability and the $SU(2)$ symmetry.  In our case, they should be fixed by the configuration of the D-branes.  Indeed, we can see from (\ref{admSU(2)}) that the defect interaction in the dCFT corresponds to choosing $\xi_{\pm}=1/2$ in the open $SU(2)$ Hamiltonian (\ref{SU(2)-Hamiltonian}), in which case the boundary $S$-matrices in (\ref{SU(2)BE}) become trivial.  

In the next section, we will derive these results with the coordinate Bethe ansatz approach, where the Bethe wavefunction plays a central role.  The analysis of the wavefunction helps us to understand how the boundary conditions of an open string is encoded into the spin chain.

\section{Coordinate Bethe Ansatz}

In this section we consider the Bethe ansatz equations for the open Heisenberg XXX${}_{\pm 1/2}$ chain by the coordinate Bethe ansatz approach.  Here XXX${}_{1/2}$ chain corresponds to the integrable spin chain model for the $SU(2)$ sector, while XXX${}_{-1/2}$ to the $SL(2)$ sector.  In the previous section, our one-loop analysis is restricted to the $SU(2)$ sector. It is basically because the direct perturbative calculation for the $SL(2)$ sector of the dCFT is technically difficult.  
In this section, however, the analysis of the $SL(2)$ sector will be included with some assumption on the boundary terms, which is deduced from the boundary condition for the corresponding open string.

\subsection{Open ${SU(2)}$ Chain}

As we previously mentioned, the boundary coefficients of the open $SU(2)$ spin chain cannot be fixed only by assuming the symmetry and the integrability.  In this section we pay a special attention to the open spin chain Hamiltonian of the form\footnote{\, In this section we will omit the prefactor $\lambda/(8\pi^{2})$ in front of $H_{{\rm XXX}_{1/2}}^{\rm open}$, since it is not essential to our arguments. }
\begin{equation}\label{open-SU(2)-Hamiltonian}
H_{{\rm XXX}_{1/2}}^{\rm open}=\sum_{l=1}^{L-1}\ko{I_{l,l+1}-P_{l,l+1}} +C_{1}\ko{I-Q_{1}^{W}}+C_{L}\ko{I-Q_{L}^{W}}\,.
\end{equation}
As we have seen in the previous section, the $SU(2)$ sector of the dCFT has the boundary coefficients $C_{1}=C_{L}=2$, but it would be useful not to fix them for the moment and argue the boundary effects in general.

Following \cite{Alcaraz}, we will show how to derive Bethe ansatz equations for the open Heisenberg Hamiltonian (\ref{open-SU(2)-Hamiltonian}).

\subsubsection*{\bmt{M=1} Case}

First we consider the single-magnon case.  For the magnon at site $x$, the eigenvalue equation becomes
\begin{equation}
H|\Psi^{(1)}\rangle=E^{(1)}|\Psi^{(1)}\rangle\,,
\end{equation}
with the Hamiltonian (\ref{open-SU(2)-Hamiltonian}) and the trial eigenfunction
\begin{align}
&\atopfrac{\hspace{3.65cm}\atopfrac{x}{\downarrow}}
{|\Psi^{(1)}(p)\rangle=\sum\limits_{1\leq x\leq L} \psi(x)\, 
\ket{Z\dots Z W Z\dots Z}\,,}\\
&\qquad \mbox{with}\quad \psi (x)=A(p)e^{ipx}-A(-p)e^{-ipx}\,.\label{f1}
\end{align}
The condition that the wavefunction $|\Psi^{(1)}\rangle$ is an eigenstate of $H$ translates to:
\begin{equation}\label{bulk condition}
E^{(1)}\psi (x)=2\psi (x)-\psi (x-1)-\psi (x+1)\,,
\qquad x=2,\dots,L-1\,,
\end{equation}
and the boundary conditions are
\begin{alignat}{3}
E^{(1)}\psi (1)&=\ko{1+C_{1}}\psi (1)-\psi (2)\,,&\qquad &\mbox{for}\quad x=1\,,\label{bc1}\\
E^{(1)}\psi (L)&=\ko{1+C_{1}}\psi (L)-\psi (L-1)\,,&\qquad &\mbox{for}\quad x=L\,.\label{bc2}
\end{alignat}
Requiring that the condition (\ref{bulk condition}) also holds for the boundaries $x=1$ and $x=L$, (\ref{bc1}) and (\ref{bc2}) reduce to the following conditions: 
\begin{equation}
\psi (0)=\ko{1-C_{1}}\psi (1)\,,\qquad \psi (L+1)=\ko{1-C_{L}}\psi (L)\,.\label{bc}
\end{equation}
Plugging (\ref{f1}) with (\ref{bc}), we obtain
\begin{equation}\label{moment-cond}
1=e^{2ipL}B_{1}(-p)B_{L}(p)
\end{equation}
with the boundary $S$-matrices,
\begin{align}
B_{1}(p)&\eq \f{e^{-ip/2}A(-p)}{e^{ip/2}A(p)}=\f{e^{-ip}-\ko{1-C_{1}}}{1-\ko{1-C_{1}}e^{-ip}}\,,\label{B1}\\
B_{L}(p)&\eq \f{e^{-ip\ko{\f{1}{2}+L}}A(-p)}{e^{ip\ko{\f{1}{2}+L}}A(p)}=\f{e^{ip}-\ko{1-C_{L}}}{1-\ko{1-C_{L}}e^{ip}}\label{BL}\,.
\end{align}
The factors $e^{\pm ip/2}$ and $e^{\pm ip\ko{\f{1}{2}+L}}$ in (\ref{B1}) and (\ref{BL}) are necessary to take account of the correct identification of the endpoints as we mentioned just below (\ref{CPO}).  
The energy can be obtained by plugging the ansatz (\ref{f1}) with the eigenvalue equation (\ref{bulk condition}), we have
\begin{equation}
E^{(1)}=4\sin^{2}\ko{\f{p}{2}}\,.
\end{equation}
It can also be written in terms of the rapidity $u$ defined by
\begin{equation}
u\equiv \f{1}{2}\cot\f{p}{2}\,,\label{u<->p}
\end{equation}
in which case we have
\begin{equation}
E^{(1)}=\f{1}{u^{2}+\f{1}{4}}\,.
\end{equation}
Note that the energy does not depend on the boundary coefficients.

One can determine the wavefunction $\psi (x;p)$ up to a factor that is invariant by the negation of the quasi-momentum $p\leftrightarrow -p$.  We have 
\begin{align}
\psi (x;p)&=a(p)\kkko{\kko{1-B_{1}(p)}\cos\ko{p\ko{x-1/2}}+i\kko{1+B_{1}(p)}\sin\ko{p\ko{x-1/2}}}\,,\\
&=b(p)\kkko{\kko{1-B_{L}(p)}\cos\ko{p\ko{x-L-1/2}}+i\kko{1+B_{L}(p)}\sin\ko{p\ko{x-L-1/2}}}
\end{align}
in terms of $B_{1}(p)$ and of $B_{L}(p)$, respectively.  Here $a(p)$ and $b(p)$ are some normalization constants.  It can be readily seen that at boundaries the wavefunction satisfies the Dirichlet boundary condition when $B_{1}=B_{L}=1$, and Neumann boundary condition when $B_{1}=B_{L}=-1$.  In both cases the condition (\ref{moment-cond}) reduces to
\begin{equation}
1=e^{2ipL}\quad \Rightarrow\quad p=\f{\pi n}{L}\qquad \ko{n\in \mathbb Z}\,.
\end{equation}
Turning to our actual case where $C_{1}=C_{L}=2$, the boundary $S$-matrices are $B_{1}=B_{L}=1$, hence the wavefunction of our concern is determined as
\begin{align}
\atopfrac{\hspace{6.95cm}\atopfrac{x}{\downarrow}}
{|\Psi^{(1)}(p)\rangle_{{SU(2)}}=a_{n}\sum\limits_{1\leq x\leq L} \sin\kko{\f{\pi n}{L}\ko{x-\f{1}{2}}}\ket{Z\dots Z W Z\dots Z}}
\end{align}

\subsubsection*{\bmt{M=2} Case}

Next we turn to the two-magnon case,
\begin{equation}
H|\Psi^{(2)}\rangle=E^{(2)}|\Psi^{(2)}\rangle\,.
\end{equation}
The ansatz for the wavefunction is
\begin{align}
&\atopfrac{\hspace{4.53cm}\atopfrac{x_1}{\downarrow}\hspace{1.18cm}\atopfrac{x_2}{\downarrow}}
{|\Psi^{(2)}\rangle=\sum\limits_{1\leq x_1 < x_2\leq L} \psi(x_1,x_2)\, 
\ket{Z\dots Z W Z \dots Z W Z\dots Z}\,,}\\
&\qquad \mbox{with}\quad \psi (x_{1},x_{2})=
A(p_{1},p_{2})e^{i\ko{p_{1}x_{1}+p_{2}x_{2}}}
-A(-p_{1},p_{2})e^{-i\ko{p_{1}x_{1}-p_{2}x_{2}}}\cr
&\hspace{4.0cm}{}-A(p_{1},-p_{2})e^{i\ko{p_{1}x_{1}-p_{2}x_{2}}}
+A(-p_{1},-p_{2})e^{-i\ko{p_{1}x_{1}+p_{2}x_{2}}}\cr
&\hspace{4.0cm}{}-A(p_{2},p_{1})e^{i\ko{p_{2}x_{1}+p_{1}x_{2}}}
+A(p_{2},-p_{1})e^{i\ko{p_{2}x_{1}-p_{1}x_{2}}}\cr
&\hspace{4.0cm}{}+A(-p_{2},p_{1})e^{-i\ko{p_{2}x_{1}-p_{1}x_{2}}}
-A(-p_{2},-p_{1})e^{-i\ko{p_{2}x_{1}+p_{1}x_{2}}}\,.\label{f2}
\end{align}
Let us first consider the bulk conditions.  Here whether the two magnons are nearest neighbors or not is crucial.  They read:
\begin{alignat}{3}
\mbox{for\quad $x_{2}>x_{1}+1$,}&\qquad &
E^{(2)}\psi (x_{1},x_{2})&=4\psi (x_{1},x_{2})-\psi (x_{1}-1,x_{2})-\psi (x_{1}+1,x_{2})-{}\cr
&{}&{}&\qquad{}-\psi (x_{1},x_{2}-1)-\psi (x_{1},x_{2}+1)\,,\label{bulk}\\
\mbox{for\quad $x_{2}=x_{1}+1$,}&\qquad &
E^{(2)}\psi (x_{1},x_{2})&=2\psi (x_{1},x_{2})-\psi (x_{1}-1,x_{2})-\psi (x_{1},x_{2}+1)\,.\label{bulk2}
\end{alignat}
From these conditions we get the following consistency condition (known as `meeting condition'),
\begin{equation}
0=2\psi (x_{1},x_{1}+1)-\psi (x_{1},x_{1})-\psi (x_{1},x_{1})\,,\label{meeting}
\end{equation}
with the dispersion relation
\begin{equation}\label{dispersion}
E^{(2)}=4\sin^{2}\ko{\f{p_{1}}{2}}+4\sin^{2}\ko{\f{p_{2}}{2}}\,.
\end{equation}
In addition to these conditions, we also have the boundary conditions
\begin{equation}
\psi (0,x_{2})=\ko{1-C_{1}}\psi (1,x_{2})\,,\qquad \psi (x_{1},L+1)=\ko{1-C_{L}}\psi (x_{1},L)\,.\label{bc-2}
\end{equation}
Then from the condition (\ref{meeting}) with the ansatz (\ref{f2}), we have the following necessary conditions:
\begin{equation}\label{bulk-cond}
 \f{A(p_{1},p_{2})}{A(p_{2},p_{1})}=\f{2-e^{-ip_{1}}-e^{ip_{2}}}{2-e^{ip_{1}}-e^{-ip_{2}}}\,,
\end{equation}
giving four conditions in total, corresponding to all the possible permutations and negations.  
On the other hand, from the boundary condition (\ref{bc-2}), we have $4+4$ conditions,
\begin{equation}\label{boundary-cond}
\f{A(-p_{1},p_{2})}{A(p_{1},p_{2})}=\f{1-\ko{1-C_{1}}e^{ip_{1}}}{1-\ko{1-C_{1}}e^{-ip_{1}}}\,,\qquad 
\f{A(p_{1},-p_{2})}{A(p_{1},p_{2})}=\f{1-\ko{1-C_{L}}e^{-ip_{2}}}{1-\ko{1-C_{L}}e^{ip_{2}}}\, e^{2i\ko{L+1}p_{2}}\,.
\end{equation}
Collecting all these twelve conditions, we arrive at the following compatibility conditions:
\begin{align}
e^{2ip_{1}L}&=B_{1}(-p_{1})B_{L}(p_{1})S_{SU(2)}\ko{p_{1},p_{2}}S_{SU(2)}\ko{p_{1},-p_{2}}\,,\\
\mbox{and}\qquad 
e^{2ip_{2}L}&=B_{1}(-p_{2})B_{L}(p_{2})S_{SU(2)}\ko{p_{2},p_{1}}S_{SU(2)}\ko{p_{2},-p_{1}}\,,
\end{align}
where, as before, $B_{1}\ko{p}$ and $B_{L}\ko{p}$ are the boundary $S$-matrices defined at the endpoints, and $S_{SU(2)}(p)$ is the bulk $S$-matrix defined by the negation of l.h.s.\ of (\ref{bulk-cond}):
\begin{equation}\label{S-SU(2)}
S_{SU(2)}\ko{p_{1},p_{2}}=-\f{2-e^{-ip_{1}}-e^{ip_{2}}}{2-e^{ip_{1}}-e^{-ip_{2}}}\,.
\end{equation}
These conditions can be rewritten in terms of the rapidities $u_{1}$, $u_{2}$ as
\begin{align}
&\ko{\f{u_{1}+i/2}{u_{1}-i/2}}^{2L}=\f{2C_{1}u_{1}+\ko{2-C_{1}}i}{2C_{1}u_{1}-\ko{2-C_{1}}i}\cdot
\f{2C_{L}u_{1}+\ko{2-C_{L}}i}{2C_{L}u_{1}-\ko{2-C_{L}}i}\cdot
\f{u_{1}-u_{2}+i}{u_{1}-u_{2}-i}\cdot
\f{u_{1}+u_{2}+i}{u_{1}+u_{2}-i}\,,\\
&\ko{\f{u_{2}+i/2}{u_{2}-i/2}}^{2L}=\f{2C_{1}u_{2}+\ko{2-C_{1}}i}{2C_{1}u_{2}-\ko{2-C_{1}}i}\cdot
\f{2C_{L}u_{2}+\ko{2-C_{L}}i}{2C_{L}u_{2}-\ko{2-C_{L}}i}\cdot
\f{u_{2}-u_{1}+i}{u_{2}-u_{1}-i}\cdot
\f{u_{2}+u_{1}+i}{u_{2}+u_{1}-i}\,.
\end{align}
These are the Bethe ansatz equations for the open Heisenberg chain with Hamiltonian (\ref{open-SU(2)-Hamiltonian}) with two magnons.  
We can determine the coefficient $A(p_{1},p_{2})$ from the conditions (\ref{bulk-cond})-(\ref{boundary-cond}) up to a factor which is invariant under the interchange of two momenta and the negation of either of the momenta.  One finds that
\begin{align}
A(p_{1},p_{2})&=e^{-i\ko{L+1}\ko{p_{1}+p_{2}}}\ko{1-\ko{1-C_{L}}e^{ip_{1}}}\ko{1-\ko{1-C_{L}}e^{ip_{2}}}\times{}\cr
&\quad \times{}e^{-ip_{2}}\ko{1-2e^{ip_{1}}+e^{i\ko{p_{1}+p_{2}}}}\ko{1-2e^{ip_{2}}+e^{i\ko{-p_{1}+p_{2}}}}
\end{align}
solves the twelve conditions and the Bethe ansatz equations.

\subsubsection*{General Case}
We extend the argument to the general $M$-magnon case, where the Bethe wavefunction is given by
\begin{equation}
\psi\ko{x_{1},\dots, x_{M}}=\sum_{\rm P, N}\epsilon_{\rm P, N} \, A\ko{p_{1},\dots ,p_{M}}\, e^{i\ko{p_{1}x_{1}+\dots +p_{M}x_{M}}}\,,
\end{equation}
where the sum is taken over all the permutations and the negations of $\{ p_{1},\dots, p_{M} \}$, and for each of such mutations, $\epsilon_{\rm P, N}$ changes sign.  The coefficients are found to be
\begin{align}
A\ko{p_{1},\dots ,p_{M}}&=\prod_{j=1}^{M}\, e^{-i\ko{L+1}p_{j}}\ko{1-\ko{1-C_{L}}e^{ip_{j}}}\times{}\cr
&\quad {}\times \prod_{1\leq j<k\leq M} e^{-ip_{k}}\ko{1-2e^{ip_{j}}+e^{i\ko{p_{j}+p_{k}}}}\ko{1-2e^{ip_{k}}+e^{i\ko{-p_{j}+p_{k}}}}
\,.\label{coeff-M}
\end{align}
The quasi-momenta $\{ p_{j} \}$ satisfy the following Bethe ansatz equation:
\begin{equation}
e^{2ip_{j}L}=B_{1}(-p_{j})B_{L}(p_{j})\,\prod_{k=1,k\neq j}^{M}S_{SU(2)}(p_{j},p_{k})S_{SU(2)}(-p_{j},p_{k})\,,
\end{equation}
or in terms of the rapidities,
\begin{align}\label{BAE-M}
\ko{\f{u_{j}+i/2}{u_{j}-i/2}}^{2L}&=\f{2C_{1}u_{j}+\ko{2-C_{1}}i}{2C_{1}u_{j}-\ko{2-C_{1}}i}\cdot
\f{2C_{L}u_{j}+\ko{2-C_{L}}i}{2C_{L}u_{j}-\ko{2-C_{L}}i}\times {}\cr
&\quad {}\times \prod_{k=1,k\neq j}^{M}\ko{\f{u_{j}-u_{k}+i}{u_{j}-u_{k}-i}}\ko{\f{u_{j}+u_{k}+i}{u_{j}+u_{k}-i}}\,,
\end{align}
for $j=1,\dots, M$.  The energy of the $M$-magnon state is given by the sum of that of each magnon state,
\begin{equation}\label{AD-SU(2)}
E^{(M)}=2\sum_{j=1}^{M}\ko{1-\cos k_{j}}=\sum_{j=1}^{M}\f{1}{u_{j}^{2}+\f{1}{4}}\,,
\end{equation}
which is still independent of the boundary terms $C_{1,L}$.

\subsection{Open ${SL(2)}$ Chain}

In the closed chain case, the coordinate Bethe ansatz for the $SL(2)$ sector has been studied by Staudacher in \cite{factorized}, where the other two-spin sectors, $SU(2)$ and $SU(1|1)$, are also studied.  We will briefly review the idea below.

The closed spin chain Hamiltonian in the $SL(2)$ sector of {\Nf} SYM has been obtained in \cite{B} and is given by
\begin{equation}
H_{{\rm XXX}_{-1/2}}=\sum_{l=1}^{L}\H^{l,l+1}_{{\rm XXX}_{-1/2}}\,.
\end{equation}
The Hamiltonian density acts on the state vector $\ket{\dots k_{l},k_{l+1},\dots}$ in the following way:
\begin{equation}\label{SL(2)-Hamiltonian}
\H^{l,l+1}_{{\rm XXX}_{-1/2}}\ket{n,n-k}=\sum_{k'=0}^{n}\ko{\delta_{k=k'}\ko{h(k)+h(n-k)}-\f{\delta_{k\neq k'}}{\left| k-k' \right|}}|k',n-k'\rangle\,,
\end{equation}
where $h(k)$ are harmonic numbers defined by $h(k)=\sum_{j=1}^{k}1/j$, and $\delta_{k=k'}\eq \delta_{k,k'}$ and $\delta_{k\neq k'}\eq 1-\delta_{k,k'}$ in terms of the standard Kronecker delta.  

As we are interested in the open spin chain Hamiltonian for this sector, we need to calculate the boundary terms associated with the defect interaction in the $SL(2)$ sector by using the dCFT, where the operator with the magnon number $S$ takes the form
\begin{equation}
 \mathcal{O} = \overline{q}_1 \ko{D^{S_{1}}Z}\dots\ko{D^{S_{L}}Z} q_2 + \cdots\,,\qquad S\eq S_{1}+\dots +S_{L}\,,\qquad D\eq D_{0}+iD_{1}\,.
\label{SL(2)op}
\end{equation}
In the $SU(2)$ case, we have determined them through the perturbative calculation of one-loop diagrams in Section \ref{sec:one-loop-SU(2)}, and then derived the Bethe ansatz equations for the open spin chain.  In the $SL(2)$ sector, however, we have not yet succeeded in performing a direct perturbative calculation in the dCFT of our concern, such as
\begin{equation}
\left\langle \kko{\overline{q}_1D Z\ko{0}} \overline{Z}\ko{z_{2}}q_{1}\ko{y_{1}}\right\rangle\,.
\end{equation}
Nevertheless we can expect that the resultant contribution of the defect interaction vanishes from the following argument. 
According to our ansatz (\ref{ansatz-SL(2)}), the string solution in the $SL(2)$ sector cannot obey the Dirichlet boundary condition, since the ${\rm AdS}_{3}\times {\rm S}^{1}$-subspace, in which the whole of the open string of this sector is immersed, is completely contained in the {\AdSs}-brane where the endpoints of the open string can freely move.  As we know from the well-established scenario for other bosonic sectors \cite{DM,EM}, in light of AdS/CFT duality, the excitations propagating on the open string can be naturally mapped to the magnon wave states on the corresponding open spin chain.  These observations lead us to argue that in our $SL(2)$ sector of the dCFT have vanishing boundary terms so that the endpoints also subject to the Neumann boundary condition.  \\



Thus it would be sufficient to consider an open $SL(2)$ spin chain Hamiltonian with free boundary (no boundary terms), 
which takes the same form as the closed case (\textit{without} periodicity),
\begin{equation}
H_{{\rm XXX}_{-1/2}}^{\rm open}=\sum_{l=1}^{L-1}\H^{l,l+1}_{{\rm XXX}_{-1/2}}\,,
\end{equation}
only to reduce the number of sites by one.
The Bethe wavefunction for the single-impurity case is given by
\begin{equation}
\atopfrac{\hspace{4.5cm}\atopfrac{x}{\downarrow}}
{|\Psi^{(1)}(p)\rangle_{{SL(2)}}=\sum\limits_{1\leq x\leq L} \psi(x)\, 
\ket{Z\dots Z \ko{DZ} Z\dots Z}\,,}
\end{equation}
and for the two-impurity case, 
\begin{align}\label{BetheStateSL}
&\atopfrac{\hspace{6.6cm}\atopfrac{x_1}{\downarrow}\hspace{1.65cm}\atopfrac{x_2}{\downarrow}}
{|\Psi^{(2)}(p_{1},p_{2})\rangle_{SL(2)}=\sum\limits_{1\leq x_1 \leq x_2\leq L} \psi(x_1,x_2)\, 
\ket{Z\dots Z (DZ) Z \dots Z (DZ) Z\dots Z}\,,}
\end{align}
where $\psi (x)$ and $\psi (x_{1},x_{2})$ are defined as (\ref{f1}) and (\ref{f2}), respectively.  Note that the two impurities can occupy the same site in this case.  
We can read off from (\ref{SL(2)-Hamiltonian}) the action of the Hamiltonian density on the neighboring spin states:
\begin{align}
&\komoji{\H\ket{\dots\ko{DZ} Z\dots }=\ket{\dots\ko{DZ} Z\dots }-\ket{\dots Z\ko{DZ}\dots }}\,,\cr
&\komoji{\H\ket{\dots Z\ko{DZ}\dots }=-\ket{\dots\ko{DZ} Z\dots }+\ket{\dots Z\ko{DZ}\dots }}\,,\cr
&\komoji{\H\ket{\dots \ko{D^{2}Z}Z\dots }=\f{3}{2}\ket{\dots \ko{D^{2}Z}Z\dots }-\ket{\dots \ko{DZ}\ko{DZ}\dots }-\f{1}{2}\ket{\dots Z\ko{D^{2}Z}\dots }}\,,\cr
&\komoji{\H\ket{\dots \ko{DZ}\ko{DZ}\dots }=-\ket{\dots \ko{D^{2}Z}Z\dots }+2\ket{\dots \ko{DZ}\ko{DZ}\dots }-\ket{\dots Z\ko{D^{2}Z}\dots }}\,,\cr
&\komoji{\H\ket{\dots Z\ko{D^{2}Z}\dots }=-\f{1}{2}\ket{\dots \ko{D^{2}Z}Z\dots }-\ket{\dots \ko{DZ}\ko{DZ}\dots }+\f{3}{2}\ket{\dots Z\ko{D^{2}Z}\dots }}\,,\cr
&\komoji{\H\ket{\dots \ko{D^{3}Z}Z\dots }=\f{11}{6}\ket{\dots \ko{D^{3}Z}Z\dots }-\ket{\dots \ko{D^{2}Z}\ko{DZ}\dots }-{}}\cr
&\hspace{4.0cm}\komoji{{}-\f{1}{2}\ket{\dots \ko{DZ}\ko{D^{2}Z}\dots }-\f{1}{3}\ket{\dots Z\ko{D^{3}Z}\dots }}\,,\nonumber
\end{align}
and so on.  Thus we have
\begin{alignat}{3}
\mbox{for\quad $x_{2}>x_{1}$,}&\qquad &
E^{(2)}\psi (x_{1},x_{2})&=4\psi (x_{1},x_{2})-\psi (x_{1}-1,x_{2})-\psi (x_{1}+1,x_{2})-{}\cr
&{}&{}&\qquad{}-\psi (x_{1},x_{2}-1)-\psi (x_{1},x_{2}+1)\,,\\
\mbox{for\quad $x_{2}=x_{1}$,}&\qquad &
E^{(2)}\psi (x_{1},x_{2})&=3\psi (x_{1},x_{2})-\psi (x_{1}-1,x_{2})-\psi (x_{1},x_{2}+1)-{}\cr
&{}&{}&\qquad{}-\f{1}{2}\psi (x_{1}-1,x_{2}-1)-\f{1}{2}\psi (x_{1}+1,x_{2}+1)\,.
\end{alignat}
and the `meeting condition' becomes,
\begin{equation}
0=2\psi (x_{1},x_{1})-2\psi (x_{1}+1,x_{1})-2\psi (x_{1},x_{1}-1)+\psi (x_{1}-1,x_{1}-1)+\psi (x_{1}+1,x_{1}+1)\,,
\end{equation}
with the same dispersion relation as (\ref{dispersion}).  
From this condition we get the Bethe ansatz equations for the $SL(2)$ sector as
\begin{equation}
e^{2ip_{1}L}=S_{SL(2)}\ko{p_{1},p_{2}}S_{SL(2)}\ko{p_{1},-p_{2}}
\quad 
\mbox{and}
\quad 
e^{2ip_{2}L}=S_{SL(2)}\ko{p_{2},p_{1}}S_{SL(2)}\ko{p_{2},-p_{1}}\,,
\end{equation}
with the bulk $S$-matrix
\begin{equation}\label{S-SL(2)}
S_{SL(2)}\ko{p_{1},p_{2}}=-\f{2-e^{-ip_{2}}-e^{ip_{1}}}{2-e^{ip_{2}}-e^{-ip_{1}}}\,,
\end{equation}
which is just the reciprocal of that in the $SU(2)$ sector (\ref{S-SU(2)}).  As we have done for the $SU(2)$ sector, we can rewrite these equations in terms of the rapidities $\{u_{i}\}$, which take real values for the $SL(2)$ sector. In the general $M$-magmon case (where now $M$ is replaced by $S$ in (\ref{SL(2)op})), the energy is given by (\ref{AD-SU(2)}) just as in the $SU(2)$ case. 

We conclude that the bulk $S$-matrix for the $SL(2)$ sector is given by (\ref{S-SL(2)}) and the boundary $S$-matrix is expected to equal $-1$ on both sides of the chain, ensuring the Neumann boundary conditions.

\section{Scaling Limit of Open Heisenberg Chain}

So far we have explored the finite $L$ case.  
Now let us take the scaling limit.  Then the problem of solving the Bethe ansatz equation can be translated into solving the associated Riemann-Hilbert problem.   We are mainly interested in the solutions which can be obtained by `opening' the two-cut solutions, which have been well studied in the closed spin chain case \cite{BMSZ,BFST,KMMZ,KZ}.  For these solutions, the filling fraction and the anomalous dimension are expressed by means of the complete elliptic integrals. Among them, so-called `double contour' solutions in the $SU(2)$ and the $SL(2)$ sector, and `imaginary root' solutions in the $SU(2)$ sector are important.  They are known to be the gauge duals of the elliptic folded and the elliptic circular strings, respectively.

In this section we examine the elliptic solutions of those kinds in the open spin chain case.  First we take the scaling limit and construct the Bethe string solutions for the open Bethe equation (\ref{BAE-M}) by formulating the `doubling trick' in the spin chain language.  Then we show the large $L$ expanded expression of the anomalous dimensions of the open operators (\ref{SU(2)op}), for each solution in turn.  We also discuss the relation between the rational (single-cut) solution and the half-filling limit of the imaginary root solution.  Finally we make some comments on the $SU(3)$ sector.

\subsection{Doubling Trick in Gauge Theory\label{sec:DT-gauge}}

Let us consider the scaling limit of the Bethe equation.  
As usual,  by rescaling $u_{j}\mapsto Lx_{j}$ after taking log,  the Bethe equation (\ref{BAE-M}) can be recast into the following form,
\begin{align}\label{resc-BAE}
\pm \f{2}{x_{j}}-2\pi n=\f{2}{L}\sum_{k=1,\,k\neq j}^{M}\ko{\f{1}{x_{j}-x_{k}}+\f{1}{x_{j}+x_{k}}}\,,
\end{align}
with arbitrary mode number $n$.  The ``$\pm$'' signature in front of the l.h.s.\ of (\ref{resc-BAE}) corresponds, respectively, to the $SU(2)$ (``$+$'') and to the $SL(2)$ (``$-$'') case.  Here the symbol $M$ denotes the `magnon' number, which is either $J_{2}$ (the number of $W$-impurity) in the $SU(2)$ case, or $S$ (the number of $D$-impurity) in the $SL(2)$ case.  Note that in both cases the contribution of the boundary $S$-matrices have been dropped in the strict scaling limit $L\to \infty$, for it scales as $1/L$.  \footnote{\, This feature can also be captured from the viewpoint of the Landau-Lifshitz sigma model approach.  When we construct a coherent state for the open $SU(2)$ chain with the Hamiltonian (\ref{open-SU(2)-Hamiltonian}), the long wave-length macroscopic spin wave `feels' not only the $L-1$ bulk sites (here $L$ is taken to be sufficiently large), but also the $1/2+1/2$ extra sites which correspond to the two defect fields $\overline{q}^{m}$, $q^{m}$.  Hence, in effect, there are $L$ sites in total, giving the same spatial occupation to the spin wave as in the closed case.}

In formulating the `doubling trick' in the $SU(2)$ and the $SL(2)$ sectors of the gauge theory, we assume that all the rescaled Bethe roots $\{ x_{j} \}$ have the same mode-number $n$, and they form a single contour in the rescaled complex plane.  Now let us add another set of $M$ Bethe roots defined by $\{x_{M+k}\}_{k=1,\dots,M}\eq \{ -x_{k}\}_{k=1,\dots,M}$.  These new Bethe roots have the mode number $-n$ and form another contour in the complex plane that is just the symmetric image of the original Bethe string.  
Then we can rewrite (\ref{resc-BAE}) as
\begin{equation}\label{closedBAE}
\pm \f{1}{x_{j}}-\pi n=\f{2}{2L}\sum_{k=1,\,k\neq j}^{2M}\f{1}{x_{j}-x_{k}}\,,
\end{equation}
which is just the Bethe ansatz equation for the closed chain whose length is now $2L$ with $2M$ impurities.  
The point is that, in order for this reinterpretation to work well, the mode number $n$ has to be an even integer rather than arbitrary one.  This feature can be compared to the fact that the folding or the winding number has to be even in the string theory side.  \\%

There are several types of the Bethe string solutions for the Bethe equation for the closed spin chain (\ref{closedBAE}).  One of the examples is the double contour solution, which has two cuts which distribute symmetrically with respect to the imaginary axis.  The cuts in the $SU(2)$ sector stretch in the imaginary direction (the upper of Fig.\,\ref{fig:DC-sol}), whereas those in the $SL(2)$ sector are known to lie completely on the real axis (the lower of Fig.\,\ref{fig:DC-sol}).  The imaginary root solution (Fig.\,\ref{fig:IR-sol}) is another example, which exists for the $SU(2)$ sector.  It has symmetric two cuts on the imaginary axis, with a `condensate' of the Bethe roots between the inner branch points of the two cuts.

To extract the open spin chain solutions from the closed ones listed above, we have only to divide the Bethe roots on the complex plane into two sets which are symmetric with respect to the origin \cite{CWW2,STY2}.  After we take the scaling limit, this operation means to extract either one of the symmetric two cuts, identifying the other as its mirror.  This corresponds, in the string theory side, to extract  only one of the modes (left/right modes) propagating on a closed string, namely the doubling trick.  As is obvious from the anomalous dimension/Bethe root relation as well as the quasi-momentum/Bethe root relation,
\begin{equation}
\gamma=\sum_{j=1}^{M}\f{1}{\f{1}{4}+u_{j}^{2}}\,,\qquad 
u_{j}=\f{1}{2}\,\cot\f{p_{j}}{2}\,,
\end{equation}
those roots symmetric about the origin have the same anomalous dimensions and can be viewed as a pair of an `incident wave' and a `reflected wave' propagating on an open string.  
Thus we have the relation
\begin{equation}\label{DT-gauge}
\gamma_{\rm (o)}(J_{1}^{\rm (o)},J_{2}^{\rm (o)})=\f{1}{2}\gamma_{\rm (c)}(J_{1}^{\rm (c)},J_{2}^{\rm (c)})\,,
\qquad J_{i}^{\rm (c)}=2J_{i}^{\rm (o)}\quad \ko{i=1,2}\,,
\end{equation}
i.e., the doubling trick for both the $SU(2)$ and the $SL(2)$ sectors holds just as in the string theory case (\ref{DT-string}).  \footnote{\, In fact, one can establish a doubling trick for the gauge theory at the level of the {\textit{un}}-rescaled Bethe roots.  In this case, the set of roots to be doubled by the trick is, one of the Bethe strings $\pm [a,b]$ for the double contour solution, and similarly, one of $\pm [is,it]$ for the imaginary root solution.}\\

It is meaningful to comment on how this doubling trick works for the various values of the boundary coefficients $C_{1,L}$ in the $SU(2)$ sector. In the case of $C_{1,L}=0$ (Neumann) or $C_{1,L}=2$ (Dirichlet), this trick works well for arbitrary $L$. (For the comparison with the string theory side, we should take $L$ large enough).  In these cases the boundary $S$-matrices reduce to $B_{1,L}=-1$ for $C_{1,L}=0$ case and $B_{1,L}=1$ for $C_{1,L}=2$ case;  they do not depend on the quasi-momenta, or spectral parameters.  
As a result, it is allowed to compare the boundary conditions in the dCFT side with those of the open spinning strings.  In fact, for the $SU(2)$ sector, we find that the boundary conditions match on both sides at finite (but large) $L$.  That is to say, the doubling trick works interestingly even 
at the finite $L$ case. \\

It is also valuable to comment on the $C_{1,L}\neq 0,2$ cases, though those are out of our present scope.  
In our setup, the D5-brane is supposed not to be a dynamical object and furthermore it does not couple to the gauge fields on the brane.  Hence the endpoints of the open string cannot satisfy the boundary conditions other than Neumann or Dirichlet one.  In the gauge theory side, on the other hand, $C_{1,L} \neq 0,2$ correspond to the cases where the endpoints of the open spin chain satisfy neither Neumann nor Dirichlet boundary condition, with rapidity-dependent phase shifts.  
Thus it would be difficult to extend our analysis to see the AdS/dCFT correspondence to the cases of $C_{1,L}\neq 0,2$.  

Nevertheless, eve in the $C_{1,L}\neq 0,2$ cases, the spin chain solutions can be a comparable object to the classical spinning string solutions, once we take the scaling limit $L\to \infty$.  In this limit, as we have mentioned, the contribution of the boundary $S$-matrices drops out of the Bethe equations written in terms of the rescaled spectral parameters.  Then we can naturally take one of the symmetric two cuts, which are the solutions of the closed spin chain, to obtain the open chain solutions.  That is, in these cases, the analysis of an open spin chain is nothing more than that of a closed spin chain, without any intrinsic feature to an open object.

\subsection{Riemann-Hilbert Problem}

In this subsection we review some basic facts about the Riemann-Hilbert problem in the closed $SU(2)$ and $SL(2)$ sector.  For details, see \cite{KMMZ,KZ}.\footnote{\, The algebraic curve for the $SO(6)$ sector ($\mathbb{R}_{t}\times {\rm S^{5}}$) has been constructed in \cite{BKSakai} and the one for the $SO(4,2)$ sector (${\rm AdS}_{5}\times {\rm S^{1}}$) is done in \cite{Schafer-Nameki:2004ik}.  This line of study has been extended further to the full $PSU(2,2|4)$ sector ({\AdSS}) in \cite{BKSZ}.}  The extension to the open cases is straightforward, since, in the scaling limit, all we have to do is to restrict the solution of closed case to the case of even $n$ (mode number), just as we have seen in the previous subsection.  \\

First note that the Bethe equation (\ref{BAE-M}) in the scaling limit can also be written in the following integral form
\begin{eqnarray}
  \pm\frac{1}{x} + 2 \pi n_l
&=&
 2 \pint_{\C_l} dx'
 \frac{\sigma(x')}{x-x'}
+
 2 \sum_{k(\neq l)} \int_{\C_k} dx'
 \frac{\sigma(x')}{x-x'},
\qquad
 x \in \C_l\,,
\label{int1}
\end{eqnarray}
where $\displaystyle \pint dx'$ means an integration for the principal value.
As before, the ``$\pm$'' signatures in front of the l.h.s.\ of (\ref{int1}) correspond, respectively, to the $SU(2)$ (``$+$'') and to the $SL(2)$ (``$-$'') case.  
The anomalous dimension (\ref{AD-SU(2)}) also translates to
\begin{eqnarray}
\gamma
=
 \frac{\lambda}{8\pi^2 L} 
 \sum_{k}
\int_{\C_k} \frac{\sigma(x)}{x^2} dx\,.
 \label{adm1i}
\end{eqnarray}
Denoting the scaling dimension of the SYM operator (\ref{SU(2)op}) or (\ref{SL(2)op}) as $\Delta$, the anomalous dimension (\ref{adm1i}) represents $\Delta-L$ in the $SU(2)$ case, and $\Delta-L-S$ in the $SL(2)$ case.

In both cases the Bethe root density for $x_j$ is defined as
\begin{eqnarray}
 \sigma(x) \equiv \frac{1}{L} \sum_{j=1}^{M} \delta(x-x_j)\,,
\end{eqnarray}
which is assumed to have the support on a set of disconnected contours $\{\C_{k}\}_{k=1,\dots,K}$ in the complex plane.  
The normalization is 
\begin{eqnarray}
 \alpha\eq \sum_{k} \int_{\C_k}  \sigma(x) dx
=\frac{M}{L}\,,
 \label{normal}
\end{eqnarray} 
where $M=J_{2}$ for the $SU(2)$ case and $M=S$ for the $SL(2)$.  
It is useful to introduce the following resolvent in solving the integral equation,
\begin{eqnarray}
 G (x) \equiv \sum_k \int_{\C_k} dx' \frac{\sigma(x')}{{x-x'}}.
\end{eqnarray}
Then the integral equation (\ref{int1}) becomes
\begin{eqnarray}
 G(x+i0) + G(x-i0) =\pm \frac{1}{x} + 2 \pi n_{l}, 
\qquad
 x \in \C_{l},
\end{eqnarray}
together with the one-loop anomalous dimension (\ref{adm1i}),
\begin{eqnarray}
  \gamma
=
 -\frac{\lambda}{8 \pi^2 L}G'(0)\,.
\label{adm1g}
\end{eqnarray}
Now introduce the Riemann surface $\Sigma$ for the general $K$-cut problem in the following form:
\begin{equation}
\Sigma:\quad y^2=x^{2K}+c_1x^{2K-1}+\dots+c_{2K-1}x+c_{2K}=\prod\limits_{k=1}^{K}\left[\ko{x-a_k}\ko{x-b_k}\right]\,,\label{Riemann}
\end{equation}
where $a_k$, $b_k$ is the endpoints of the $k$-th cut $\C_k$, endowed with the reality constraint.  
For this general elliptic curve, the resolvent can be written as 
\begin{equation}
G\left( x \right)=-{{c_{2K-1}} \over {4c_{2K}}}+x\ko{ {-{{c_{2K-2}} \over {4c_{2K}}}+{{c_{2K-1}^2} \over {16c_{2K}^2}}+{{a} \over {\sqrt {c_{2K}}}}}}+\O\left( {x^2} \right)\,,
\end{equation}
from which one can obtain the general formula for the one-loop anomalous dimension (using (\ref{adm1g})) \cite{KMMZ,KZ},
\begin{equation}
\gamma=\f{\lambda}{8\pi^2L}\ko{ {{{c_{2K-2}} \over {4c_{2K}}}-{{c_{2K-1}^2} \over {16c_{2K}^2}}-{{a} \over {\sqrt {c_{2K}}}}}}\,.\label{gen-AD}
\end{equation}
The parameter $a$ is related to the filling fraction $\al$ as
\begin{equation}
\pm\al=\f{1}{2}-a\,,
\end{equation}
where the upper sign corresponds to the $SU(2)$ case and the lower to the $SL(2)$.  
We will use this formula to discuss the general anomalous dimension formula for $K=1$ (rational) and $K=2$ (elliptic) cases in the following subsections.  

\subsection{One-loop Anomalous Dimension and Filling Fraction}

The anomalous dimension and the filling fraction of the general elliptic solution for the closed spin chain case can be seen from Appendix \ref{app:3-loop formula}, where the Riemann-Hilbert problem associated with the elliptic solutions is solved up to the three-loop level.      
Consequently, the generic formula for the one-loop anomalous dimension can be extracted from (\ref{formula2}).  The anomalous dimension of the defect operator (\ref{SU(2)op}) is related to that of a single trace operator as (\ref{DT-gauge}), and is given by
\begin{align}
\gamma_{\rm g}\ko{n;\xi}&=-\f{n^{2}\lambda}{8\pi^{2}L}\ko{\K{1-\xi}+i \K{\xi}}\cr
&\hspace{0.8cm}{}\times \kko{-2\E{1-\xi}+2i \E{\xi}+\ko{1+\xi}\K{1-\xi}-i \ko{1-\xi} \K{\xi}}\,.\label{gen_AD}
\end{align}
Here the real part of the elliptic moduli $\xi$ takes a value in $[0,1]$, and the integer $n$ is the mode number for the single (elliptic) contour associated with the open spin chain solution.  When $n$ is an even integer, say $2n'$, (\ref{gen_AD}) is equal to the one-loop anomalous dimension for the closed spin chain with mode number $n'$.
As is obvious from the `doubling trick' procedure, the filling fraction for the open spin chain is just the same as the closed spin chain case, which we see from (\ref{fraction}) as
\begin{equation}
\alpha_{\rm g}\ko{\xi}=\f{1}{2}-\f{1}{2\sqrt{\xi}}\f{\E{1-\xi}-i \ko{\E{\xi}-\K{\xi}}}{\K{1-\xi}+i \K{\xi}}\,.\label{gen_ag}
\end{equation}

\subsubsection*{ Halved Double Contour Solution}


\FIGURE{
\centerline{
\includegraphics[width=0.9\textwidth]{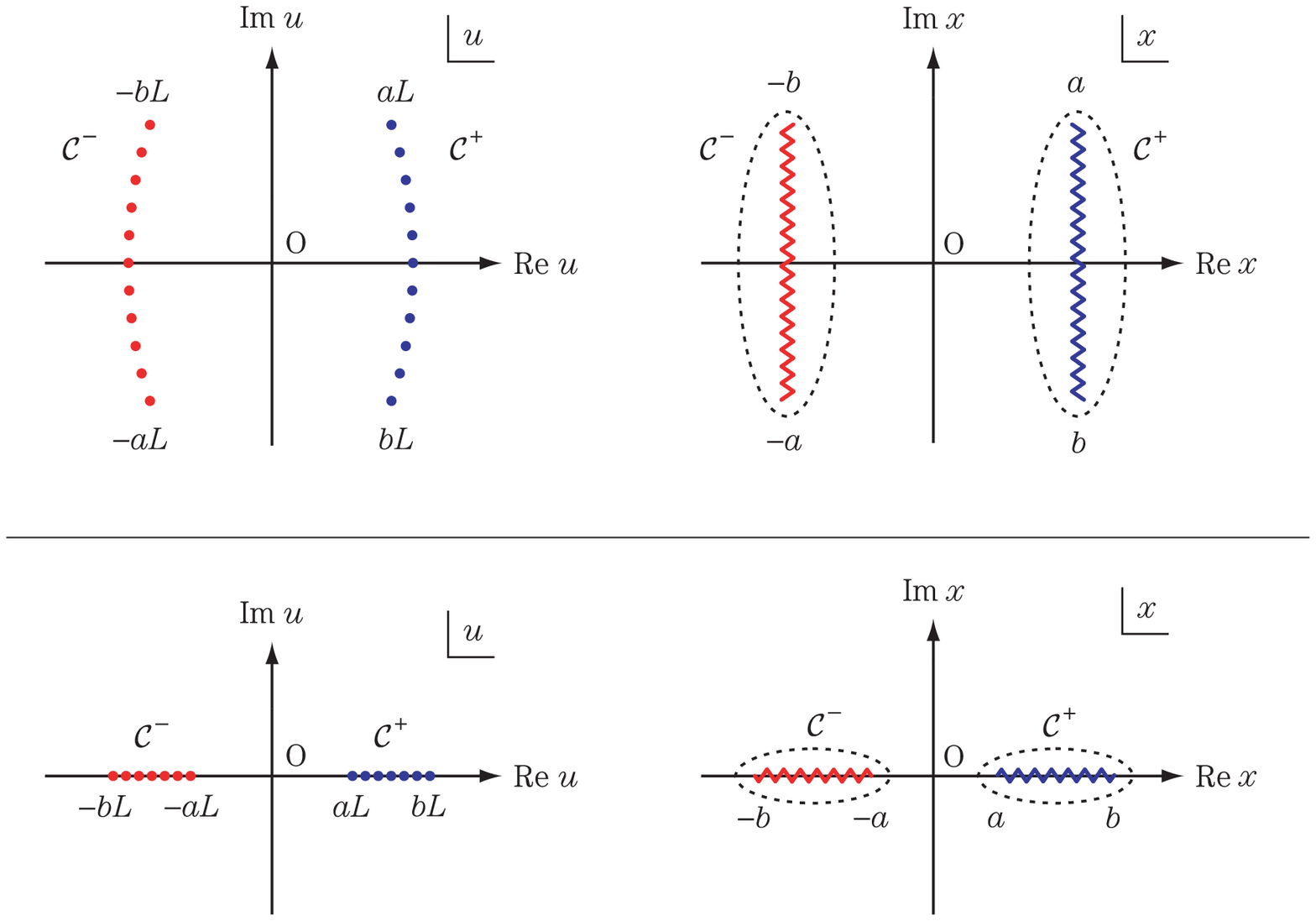}
\caption{\small (Halved) double contour solution in the $SU(2)$ (upper) and the $SL(2)$ (lower) sectors.
}
\label{fig:DC-sol}
}}

The filling fraction and the one-loop anomalous dimension of the halved double contour solution in the $SU(2)$ sector can be obtained from the generic formulae (\ref{gen_ag}) and (\ref{gen_AD}) by analytic continuing them past $\xi =0$. These are given by 
\begin{equation}
\al_{\rm DC}\ko{\xi}=\f{1}{2}-\f{1}{2\sqrt{\xi}}\f{\E{1-\xi}}{\K{1-\xi}}\,,\label{aDC-SU(2)}
\end{equation}
and 
\begin{equation}
\gamma_{\rm DC}\ko{n;\xi}=-\f{n^{2}\lambda}{8\pi^{2}L}\K{1-\xi}\kko{-2\E{1-\xi}+\ko{1+\xi}\K{1-\xi}}\,.\label{gDC-SU(2)}
\end{equation}
The halved double contour solution in the $SL(2)$ sector has almost the same expressions as (\ref{aDC-SU(2)}) and (\ref{gDC-SU(2)}), only to be negated, i.e.,
\begin{align}
\widetilde\al_{\rm DC}\ko{\zeta}&=-\f{1}{2}+\f{1}{2\sqrt{\zeta}}\f{\E{1-\zeta}}{\K{1-\zeta}}\,,\label{aDC-SL(2)}\\
\widetilde\gamma_{\rm DC}\ko{n;\zeta}&=\f{n^{2}\lambda}{8\pi^{2}L}\K{1-\zeta}\kko{-2\E{1-\zeta}+\ko{1+\zeta}\K{1-\zeta}}\,.\label{gDC-SL(2)}
\end{align}
As we have mentioned before, in order for (\ref{gDC-SU(2)}) and (\ref{gDC-SL(2)}) to represent the one-loop anomalous dimension of the open spin chain solution, which results from the closed one via the `doubling trick', the mode number $n$ has to be an even integer.  Indeed, $\gamma_{\rm DC}\ko{2n;\xi}$ and $\widetilde\gamma_{\rm DC}\ko{2n;\zeta}$ (i.e., the anomalous dimensions of the open spin chains with the mode number $2n$) are equal to those of the corresponding closed spin chains with mode number $n$, which are $n^{2}$ times Eqs.\,(2.8) and (C.12) of \cite{BFST}, respectively.

\subsubsection*{ Halved Imaginary Root Solution}


\FIGURE{
\centerline{
\includegraphics[width=0.70\textwidth]{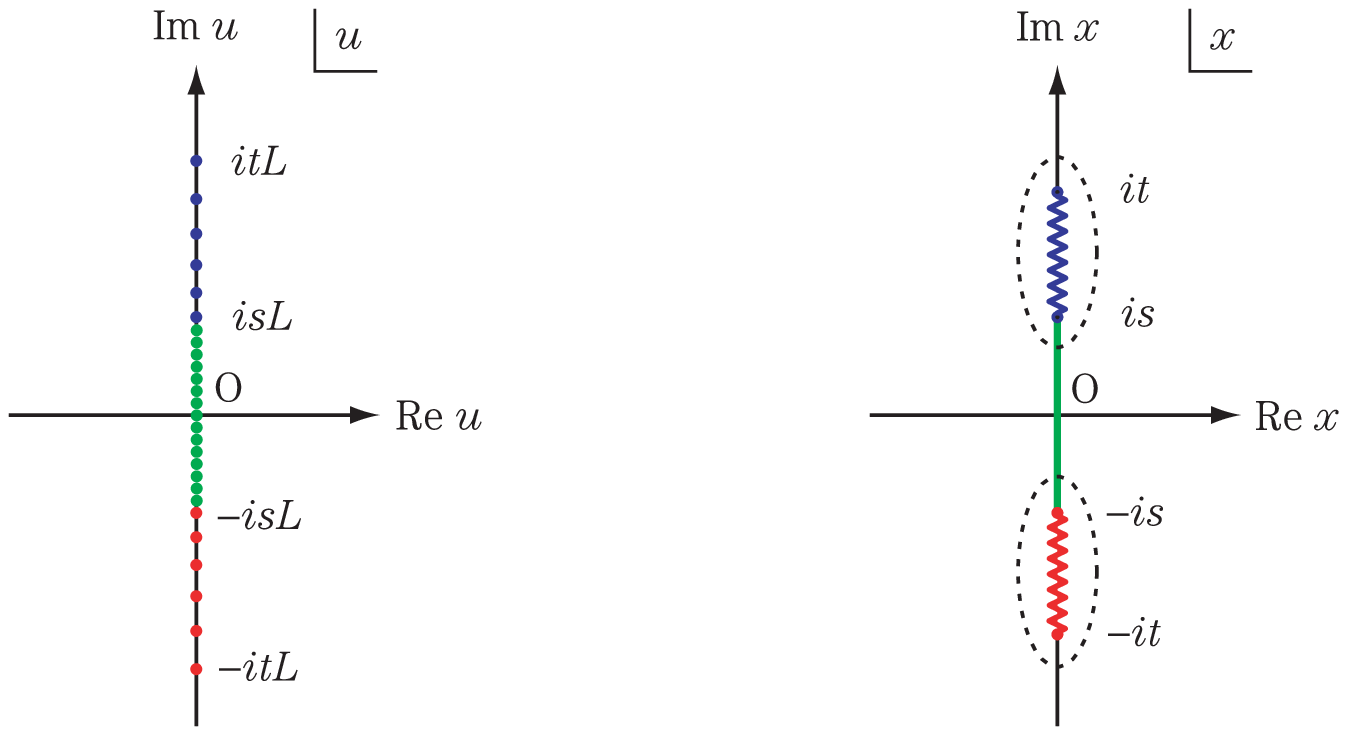}
\caption{\small (Halved) imaginary root solution in the $SU(2)$ sector.
}
\label{fig:IR-sol}
}}

The filling fraction and the one-loop anomalous dimension of the halved imaginary root solution are obtained by the following procedure.  First, let us analytic continue the generic functions (\ref{gen_ag}) and (\ref{gen_AD}) past $\xi =1$, then modular transform the resulting expressions using the relations (\ref{mod.trans-1})-(\ref{mod.trans-2}).  As the result, we obtain the filling fraction,
\begin{equation}\label{aIR}
\al_{\rm IR}\ko{\eta}=\f{\K{\eta}-\E{\eta}+\sqrt{\eta}\,\K{\eta}}{2\sqrt{\eta}\,\K{\eta}}\,,
\end{equation}
and the one-loop anomalous dimension,
\begin{equation}\label{gIR}
\gamma_{\rm IR}\ko{n;\eta}=\f{n^{2}\lambda}{8\pi^{2}L}\K{\eta}\kko{2\E{\eta}-\ko{1-\eta}\K{\eta}}\,,
\end{equation}
where the new moduli is related to the old one via $\eta=1/\xi \in [0,1]$.  
Similar to the halved double contour case, $\gamma_{\rm IR}\ko{2n;\eta}$ is equal to the one-loop anomalous dimension of closed spin chain solution having a condensate of density $n$, which is $n^{2}$ times Eq.\,(D.4) of \cite{BFST}.\footnote{\, The filling fraction $\alpha$ in Eq.\,(D.4) of \cite{BFST} is related to (\ref{aIR}) as $\alpha=1-\alpha_{\rm IR}$.}

\subsection{Endpoints of Bethe String}

It is valuable to note the expressions for the endpoints of the elliptic solutions just we have seen.
We denote the four endpoints of the elliptic solution (after taking the scaling limit) as $\pm a\ko{n;\xi}$ and $\pm b\ko{n;\xi}$ where $\xi$ is the same moduli parameter utilized to describe the filling fraction and the one-loop anomalous dimension in the previous subsection.  Their generic forms can be cast into the forms,
\begin{equation}
a\ko{n;\xi}=\f{1}{4n\ko{\K{1-\xi}+i  \K{\xi}}}\,,\qquad 
b\ko{n;\xi}=\f{1}{4n\sqrt{\xi}\ko{\K{1-\xi}+i  \K{\xi}}}\,.\label{endpt}
\end{equation}
The endpoints of the double contour solution can be obtained by analytic continuing them past $\xi =0$, and we obtain
\begin{equation}
a\ko{n;\xi}\eq \f{1}{4 n \K{1-\xi}}\,,\qquad 
b\ko{n;\xi}\eq \f{1}{4 n \sqrt{\xi}\,\K{1-\xi}}\,,\label{endpt-DC}
\end{equation}
where $0<\Re \xi <1$.  To get the endpoints of the imaginary root case, all we have to do is to rewrite the moduli from $\xi$ to $\eta=1/\xi$. As a result we have
\begin{equation}
-i t\ko{n;\eta}\eq -\f{i}{4 n \sqrt{\eta}\,\K{\eta}}\,,\qquad 
-i s\ko{n;\eta}\eq -\f{i}{4 n \K{\eta}}\,.\label{endpt-IR}
\end{equation}
Note that the elliptic moduli are related to the endpoints as
\begin{equation}\label{moduli}
\xi=\f{a\ko{n;\xi}^{2}}{b\ko{n;\xi}^{2}}\,,\qquad 
\eta=\f{s\ko{n;\eta}^{2}}{t\ko{n;\eta}^{2}}\,.
\end{equation}
Here the region $0<\Re \xi<1$ corresponds to the double contour solution, whereas $1<\Re \xi$, i.e., $0<\Re \eta<1$, to the imaginary root one.



\FIGURE{
\centerline{
\includegraphics[width=0.9\textwidth]{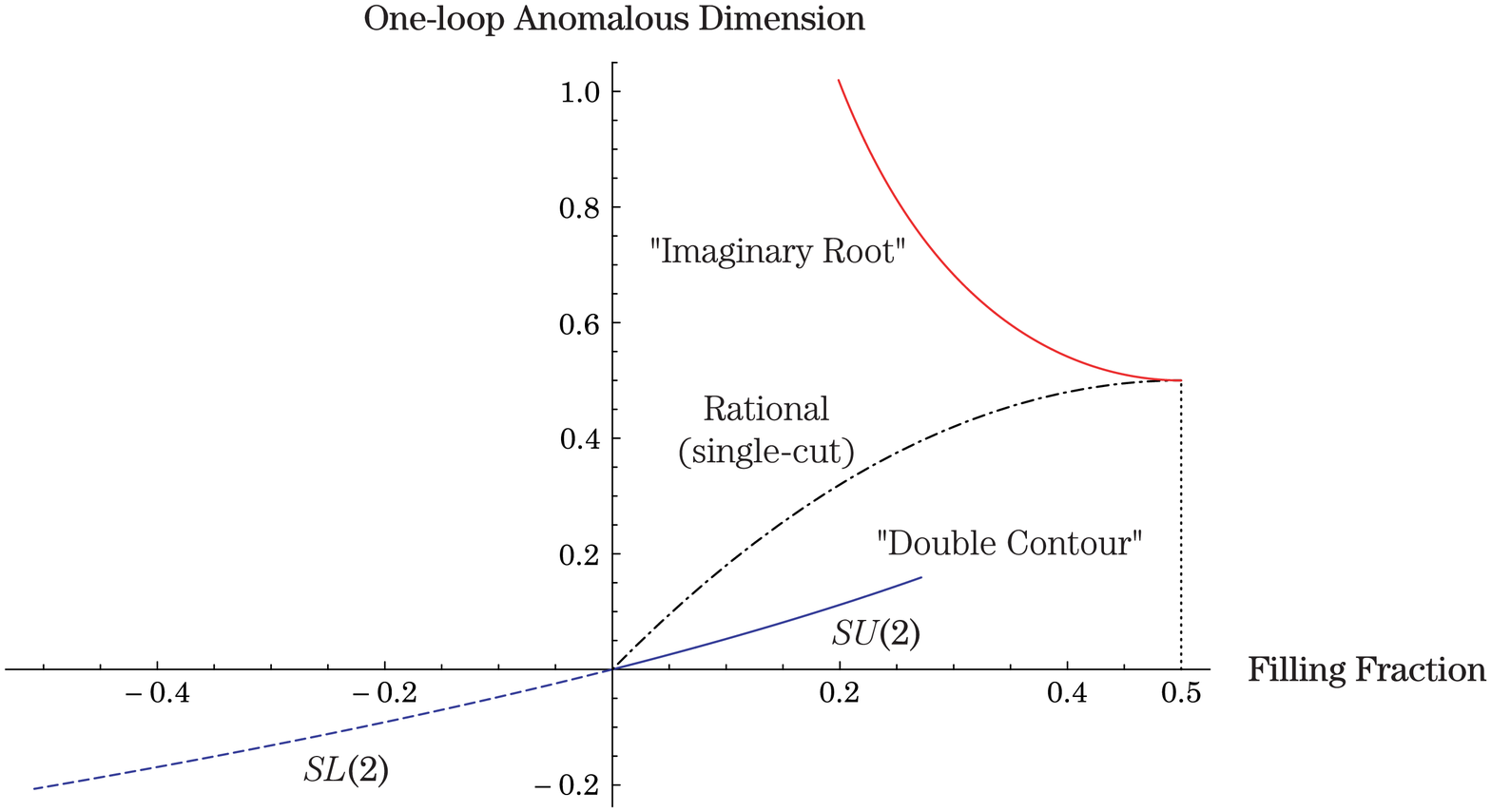}
\caption{\small One-loop anomalous dimension vs. filling fraction.  The anomalous dimension $\gamma_{\rm g}$ (divided by $\lambda/L$) for each string solution is shown as a function of $\alpha_{\rm g}$.  Here the mode number ($\B$-cycle) for the halved double contour solution is $n=2$, the condensation density ($\A$-cycle) for the halved imaginary root solution is $n=4$ and the mode number for the rational (single-cut) solution is $n=2$.   To obtain an elliptic open spin chain solution from a closed one via the doubling trick, the $\A$- or $\B$-cycle associated with the closed case has to be even.  The figure can be compared to the $\al_{\rm s}<1/2$ region of Fig.\,\ref{fig:del-vs-ros} in the string theory side.  We also note that the curve for the double contour distribution has a singular point (turning point) at $\xi=-1$.
}
\label{fig:AD-vs-fraction}
}}

\subsection{Rational Limit in Gauge Theory}

In this subsection we consider the rational solutions in the spin chain side.  We can reach a rational solution by taking $\eta\to 0$ limit in the imaginary root solution \cite{BMSZ,KMMZ}.  In this limit the outer branch points $\pm it$ in the rescaled complex plane go to infinity, whereas the inner ones remain at finite distance from the origin:
\begin{equation}
\ko{\pm is(n;\eta), \,\pm it(n;\eta)}\quad \xrightarrow[\eta\to 0]{}\quad 
\ko{\pm \f{i}{2n\pi}, \,\pm i\infty}\,.
\end{equation}
Thus the original Riemann surface with two cuts reduces to the one with single-cut, with nonzero condensate between two branch points, giving a non-vanishing $\A$-cycle for the cut.  
In this limit the filling fraction and the one-loop anomalous dimension become
\begin{equation}\label{gauge-rational}
\alpha_{\rm IR}\ko{\eta}\quad \xrightarrow[\eta\to 0]{}\quad \f{1}{2}\,,\qquad 
\gamma_{\rm IR}\ko{n;\eta}\quad \xrightarrow[\eta\to 0]{}\quad \f{n^{2}\lambda}{8L}\,.
\end{equation}
We see that they match to the string theory results (\ref{string-rational}) via the identification: $n_{\rm gauge}=2n_{\rm string}$ and $L=J$.  

This type of rational solution can be physically equivalent to another configuration of Bethe string;  a single-cut solution of the Riemann-Hilbert problem \cite{KMMZ}:  
\begin{equation}
2\pint_\C d\xi\, \f{\rho\ko{\xi}}{q-\xi}=\f{1}{q}+2\pi n\, ,\qquad \ko{q\in \C}\label{1-cut}\,,
\end{equation}
where $\C$ denotes the single Bethe string.  
In the closed spin chain case, the ratio of the momentum, which is given by
\begin{equation}
P=\f{1}{L}\sum\limits_{k=1}^J\f{1}{q-q_k}\eq m\,,
\end{equation}
to the mode number $n$ defines the filling fraction,
\begin{equation}
\alpha=\f{m}{n}\,.
\end{equation}
In terms of these parameters the elliptic curve on the Riemann surface is represented by
\begin{align}
y^2=q^2-\f{2\alpha-1}{\pi n}\, q+\f{1}{4\pi^2n^2}\,,
\end{align}
from which we can read off the coefficients defining the Riemann surface (\ref{Riemann}) as 
\begin{equation}
c_1=-\f{2\alpha-1}{\pi n},\qquad c_2=\f{1}{4\pi^2n^2}\,.
\end{equation}
Finally we get the one-loop energy for the single cut solution by inserting these coefficients to the general formula (\ref{gen-AD}) as
\begin{equation}
\gamma=\f{\lambda}{8\pi^2L}\ko{\f{1}{4r_2}-\f{r_1{}^2}{16r_2{}^2}}=\f{\lambda n^2\alpha\ko{1-\alpha}}{2L}\,.
\end{equation}
It takes the maximum value at $\alpha=1/2$, which turns out to give the same result as the $\al_{\rm IR}\to 1/2$ limit of the elliptic imaginary root case (\ref{gauge-rational}).  

The single-cut solution in the $SL(2)$ sector \cite{KZ} can be obtained just by negating the $\alpha$ and $\gamma$ simultaneously, as we have noted before, as a realization of the `analytic continuation' in gauge theory. \\

Here we should remark when and how the doubling trick works for the single-cut solution above.  
Let us remind that the Bethe roots for the closed spin chain have to distribute symmetrically with respect to the origin, in order for the doubling trick to work well.  With this in mind, we see that the doubling trick works only for the single-cut solution in the $SU(2)$ sector with $\al=1/2$.  In this case, the two branch points of the single-cut go to $\pm i/(2n\pi)$ on the imaginary axis, and the configuration of the Bethe string in the Riemann surface becomes equivalent to that of the $\al\to 1/2$ limit of the imaginary root solution, which has a well-established doubling trick interpretation between the closed and open cases.  Note that we have encountered similar situation in the string theory side, see the argument below (\ref{const-radii}).

\subsection{${SU(3)}$ sector}

We conclude this section by making some comments on the duality in the $SU(3)$ sector.  In the closed case, the $SU(3)$ sector of {\Nf} SYM is closed at one-loop level.  But, in the open case, the $SU(3)$ sector is not closed any more due to the presence of defect interactions.  To see this, let us concentrate on the left endpoint of the dCFT operator, $\overline q_1 X$, where $X$ is either one of $Z$, $W$ and $V$ with
\begin{equation}
 Z=\ko{X_{\rm H}^1 + i X_{\rm H}^2}/\sqrt{2}\,, \qquad
 W=\ko{X_{\rm V}^3 + i X_{\rm V}^4}/\sqrt{2}\,, \qquad
 V=\ko{X_{\rm H}^5 + i X_{\rm V}^6}/\sqrt{2}\,.
\end{equation}
One can show from a direct computation that the operator $\overline q_1 X$ mixes with $3\, \overline q_1 V - \overline q_1 \overline V + 2 \overline q_2 Z$ as well as $\overline q_1 Z$ and $\overline q_1 W$ from the one-loop effect. This implies that  the $SU(3)$ sector is not closed in this open case. 
This observation is consistent with the analysis in the open string side where we cannot construct three-spin solutions satisfying the boundary conditions supplied by the D5-brane.  Concretely speaking, the Cartan charge $J_{3}\eq J_{56}$ defined by (\ref{cartan}) has to be zero in our brane setup.

\section{Conclusion and Discussion} 

In the string theory side, we have considered rotating string solutions for open strings 
on an {\AdSs}-brane in the bulk {\AdSS}
background. We have first formulated the coordinate system to specify the 
sigma model charges such as energy and spins,
then we considered two-spin solution in the $SL(2)$ and the $SU(2)$
sectors with appropriate rotating string ansatze.  
On the other hand, in the gauge theory side, we computed the matrix of
anomalous dimension for the $SU(2)$ sector and showed it was
represented by an open integrable spin chain with diagonal boundary.
Then we carried out the Bethe ansatz to diagonalize the anomalous dimension 
matrix, and showed the boundary condition for the Bethe wavefunction matched 
to that of the open string in the $SU(2)$ sector of string theory.  
For the $SL(2)$ sector, we discussed 
what the contribution of the defect interaction to the 
anomalous dimension should be to satisfy the boundary condition, that is 
expected from that of corresponding string feature in light of the AdS/CFT duality.   \\


We now make some comments on the duality at the higher-loop level.  In the string theory side, the doubling trick relation (\ref{DT-string}) still holds for the higher-loop level, really for the all-loop level.  But It is not the case with the gauge theory side, where we must prove the relation (\ref{DT-gauge}) order-by-order in $\lambda$.  In section \ref{sec:one-loop-SU(2)} we computed the anomalous dimension matrix for the operators (\ref{SU(2)op}) at the one-loop level and showed it was represented by an integrable open chain Hamiltonian.  At the higher loop level, however, the integrability of the anomalous dimension matrix including the defect interactions has not been proved yet.  Hence it would be interesting to examine whether or not it supplies integrable boundaries, and if it \textit{should be} integrable, then what open spin chain model corresponds to the $SU(2)$ sector of the dCFT at the higher loop.  
In closed case, it is known that the $SU(2)$ sector can be mapped to the Inozemtsev spin chain model up to the three-loop level \cite{SS}, and it may be the case of our open spin chain with boundary conditions.  

It is also interesting to investigate how the doubling trick relations on both sides of the duality are affected by the finite-size correction. But it is not easy to compare the string energy corrected in powers of $1/J$ with the anomalous dimension of the SYM operators corrected by $1/L$, even in the closed case.\footnote{\, The one-loop correction to the energy of closed spinning strings in the $SU(2)$ and the $SL(2)$ sectors are calculated in \cite{Frolov:2004bh} and \cite{Park:2005ji} respectively.}

As another direction, it would be interesting to consider other AdS-brane cases (for the classification of $1/2$ BPS AdS-branes, see \cite{ST}-\cite{Bain:2002tq}).  The action of the dCFT is well-established in the {\AdSs} case, but otherwise it is not the case. Hence it may seem impossible to compute the matrix of anomalous dimension in the standard perturbative calculation.  
However, assuming
the integrability of the dCFT, one may obtain the expression of the anomalous
dimension matrix by the symmetry argument.  That is to say, the boundary
terms may be determined from the unbroken symmetries in the presence of the
AdS-brane.  Studies in this direction should give a promising way to clarify the AdS/dCFT duality. 

Furthermore, our formulation developed in this paper can be applied to 
the (dual) giant graviton case. In this case the anomalous dimension matrix is
represented by an open integrable spin chain as shown in \cite{BV,BCV}. 
It is interesting to consider open rotating strings on the giant graviton
according to our method.  It would be also interesting to examine a non-supersymmetric background case \cite{deMelloKoch:2005jg}, 
where the open string states (open chains) attaching to the giant graviton acquire some extra deformation parameters.  We will report on these subjects in the near future as another publication. 

We hope that our arguments for boundary conditions could be a clue to 
discuss open strings in studying the AdS/CFT duality 
at far-from BPS regime.

\acknowledgments

We would like to thank H.~Fuji, M.~Hatsuda, Y.~Hikida, K.~Ideguchi,
Y.~Imamura, V.~Kazakov, B.~MacCoy, Y.~Nakayama, K.~Sakai, M.~Staudacher, 
Y.~Susaki, D.~Tomino and A.~Yamaguchi for useful
discussion. The work of Y.~T.\ is supported in part by The 21st Century
COE Program ``Towards a New Basic Science; Depth and Synthesis.'' The
work of K.~Y.\ is supported in part by JSPS Research Fellowships for
Young Scientists.

\vspace{1.0cm}

\appendix
\section*{Appendices}

\vspace{0.5cm}

\section{Elliptic Solutions for Higher Loops\label{app:3-loop formula}}

Let us compute the anomalous dimensions of the SYM operators in the $SU(2)$ elliptic sector (for the rational case, see Appendix of \cite{Minahan:higher}).  We consider the ``long'' single trace operators composed of two complex scalars $Z$ and $W$,
\begin{equation}\label{SU(2)-closed}
\Tr\ko{Z^{J_{1}}W^{J_{2}}}+\mbox{perm.}
\end{equation}
in the limit where both $J_{1}$ and $J_{2}$ are sufficiently large (i.e., in the Frolov-Tseytlin sector).  Here $J_{2}$ is the `magnon' (or `impurity') number.  The general solution for the one-loop anomalous dimension matrix of (\ref{SU(2)-closed}) has been derived in \cite{KMMZ} from the viewpoint of the Riemann-Hilbert problem.  We extend this result to the three-loop level.

\subsection{Bethe Ansatz for Higher Loops}

Let us brush up the notion of Bethe ansatz and the associated Riemann-Hilbert problem.   In \cite{SS}, the integrable structure in the $SU(2)$ sector of SYM is identified with that of the Inozemtsev spin chain up to the three-loop level.  The asymptotic Bethe ansatz equations for the $SU(2)$ sector are given by
\begin{equation}
\ko{\f{u_{j}+i/2}{u_{j}-i /2}}^{L}=\prod\limits_{\scriptstyle k = 1 \hfill \atop 
  \scriptstyle  k \ne j \hfill}^{J_{2}} \f{\varphi_{j}-\varphi_{k}+i }{\varphi_{j}-\varphi_{k}-i }\,.
\end{equation}
The rapidities $\{ \varphi_{j} \}$ are related to the quasi-momenta $\{ p_{j} \}$ as
\footnote{\, The rapidity function $\varphi (p)$ has been conjectured to have the all-loop form
\begin{equation}
\varphi (p)=\f{1}{2}\cot\ko{\f{p}{2}}\sqrt{1+\f{\lambda}{\pi^{2}}\sin^{2}\ko{\f{p}{2}}}\nonumber
\end{equation}
for the $SU(2)$ sector.  The all-loop (asymptotic) physics has been also studied intensively \cite{BDS,factorized}.}%
\begin{equation}
\varphi\ko{p}=\f{1}{2}\cot\f{p}{2}+2\sum_{n>0}\f{t^{n}\sin p}{\ko{1-t^{n}}^{2}+4t^{n}\sin^{2}\f{p}{2}}\,,
\end{equation}
where the parameter $t=t\ko{\lambda}$ is defined by the relation
\begin{equation}
\f{\lambda}{16\pi^{2}}=\sum_{n>0}\f{t^{n}}{\ko{1-t^{n}}^{2}}\,.
\end{equation}
By using these relations, the anomalous dimensions of two typical distributions of Bethe roots, the double contour and the imaginary root solution, are calculated up to the third order in \cite{SS}.  There the Bethe equations are solved for each of the concrete distributions of the Bethe roots.  On the other hand, in the one-loop analysis, it is shown in \cite{KMMZ} that both of the solutions can be considered as the special cases of a single, general elliptic solution that has symmetric two cuts on the Riemann surfaces.  In the following, we generalize the one-loop results of \cite{KMMZ} to the three-loops.  

\subsection{Anomalous Dimension Formula at the Three-Loop Level}

We shall give a {\sl recipe} for a general two-cut, three-loop formula of 
the anomalous dimension in the $SU(2)$ sector.\footnote{\, The general two-cut problem is also discussed in \cite{Chen:2005sb} at the one-loop level.}

\subsubsection*{General Two-Cut Case}
The anomalous dimension per unit length of the spin chain, or `energy density', at the three-loop level can be computed from the resolvent $G\ko{x}$:
\begin{equation}\label{anomalous_dimension}
\f{\gamma}{L}=-\oint\f{d  x}{2\pi i }\,G\ko{x}\left(
 \f{\tlambda}{8\pi^{2}}\f{1}{x^{2}}+
  \f{3\tlambda^{2}}{128\pi^{4}}\f{1}{x^{4}}+
  \f{5\tlambda^{3}}{1024\pi^{6}}\f{1}{x^{6}}+
  \O(\tlambda^{4})
   \right)\,,
\end{equation}
where $\tlambda=\lambda/L^{2}$ is the BMN coupling constant, and the length of the spin chain is $L$.   In order to compute the energy density, all we have to do is to find out the form of $G(x)$ expanded around $x=0$.  The Riemann surface of a general two-cut solution is given by
\begin{equation}\label{Riemann_surface}
y^{2}=x^{4}+c_{1}x^{3}+c_{2}x^{2}+c_{3}x+c_{4}\,,
\end{equation}
and the two-cut solution of Bethe equations are constructed with a meromorphic differential on (\ref{Riemann_surface}).  Following the procedure proposed in \cite{KMMZ}, we introduce the three-loop level quasi-momentum,
\begin{equation}\label{quasi-momentum}
p\ko{x}=G\ko{x}-\f{1}{2x}-\f{\tlambda}{16\pi^{2}}\f{1}{x^{3}}-\f{3\tlambda^{2}}{256\pi^{4}}\f{1}{x^{5}}\,.
\end{equation}  
Here we have used the result of \cite{SS} to write down the coefficients in the r.h.s.\ of Eq.\,(\ref{quasi-momentum}).  Then the most general form of the differential is given by
\begin{equation}\label{differential}
d  p=\f{d  x}{y}\ko{\f{a_{-5}}{x^{6}}+\f{a_{-4}}{x^{5}}+\f{a_{-3}}{x^{4}}+\f{a_{-2}}{x^{3}}+\f{a_{-1}}{x^{2}}+\f{a_{0}}{x}+a_{1}}\,,
\end{equation}
and the coefficients $a_{0},\dots, a_{-5}$ are fixed by the condition that the integration of (\ref{differential}) matches up to (\ref{quasi-momentum}).
Thus we find the coefficients to be
\begin{align}
a_{0}&=\frac{3 }{32768 {\pi }^4 
      {{c_4}}^{{9/2}}}
      \left( -5 {\lambda }^2 {{c_3 ^2}} + 
        20 {\lambda }^2 {c_2} {c_4} + 
        128 {\pi }^2 \lambda  {{c_4 ^2}} \right)  
      \left( 5 {{c_3 ^3}} - 12 {c_2} {c_3} {c_4} + 
        8 {c_1} {{c_4 ^2}} \right)  \cr
&\hspace{1.0cm}{}+ 
   \frac{3  }{32768 {\pi }^4 {{c_4}}^{{9/2}}}
   \left( -3 {{c_3 ^2}} + 4 {c_2} {c_4} \right)
         \left( 5 {\lambda }^2 {{c_3 ^3}} - 
        20 {\lambda }^2 {c_2} {c_3} {c_4} + 
        40 {\lambda }^2 {c_1} {{c_4 ^2}} + 
        128 {\pi }^2 \lambda  {c_3} {{c_4 ^2}} \right) 
      \cr
&\hspace{1.0cm}{} - 
   \frac{15 {\lambda }^2 {c_3}  }{65536 {\pi }^4 
      {{c_4}}^{{9/2}}}
      \left( 35 {{c_3 ^4}} - 
        120 {c_2} {{c_3 ^2}} {c_4} + 
        96 {c_1} {c_3} {{c_4 ^2}} + 
        16 \left( 3 {{c_2 ^2}} - 4 {c_4} \right)  
         {{c_4 ^2}} \right) \cr
&\hspace{1.0cm}{}- 
   \frac{15 {\lambda }^2 }{65536 {\pi }^4 {{c_4}}^{{9/2}}}
      \left( -63 {{c_3}}^5 + 
        280 {c_2} {{c_3 ^3}} {c_4} - 
        240 {c_1} {{c_3 ^2}} {{c_4 ^2}} - 
        48 {c_3} \left( 5 {{c_2 ^2}} - 4 {c_4} \right)
            {{c_4 ^2}} + 192 {c_1} {c_2} {{c_4 ^3}}
        \right) \cr
&\hspace{1.0cm}{}
    - \frac{{c_3} }{65536 
      {\pi }^4 {{c_4}}^{{9/2}}} \left( 75 {\lambda }^2 {{c_3 ^4}} - 
        360 {\lambda }^2 {c_2} {{c_3 ^2}} {c_4} + 
        240 {\lambda }^2 {{c_2 ^2}} {{c_4 ^2}} + 
        480 {\lambda }^2 {c_1} {c_3} {{c_4 ^2}} \right.\cr
&\hspace{4.0cm}{}\left.{}+ 
        768 {\pi }^2 \lambda  {{c_3 ^2}} {{c_4 ^2}} - 
        960 {\lambda }^2 {{c_4 ^3}} - 
        3072 {\pi }^2 \lambda  {c_2} {{c_4 ^3}} - 
        16384 {\pi }^4 {{c_4 ^4}} \right)\,,\cr
a_{-1}&=  -\frac{1}{32768 
     {\pi }^4 {{c_4}}^{{7/2}}}\left( 75 {\lambda }^2 {{c_3 ^4}} - 
       360 {\lambda }^2 {c_2} {{c_3 ^2}} {c_4} + 
       240 {\lambda }^2 {{c_2 ^2}} {{c_4 ^2}} + 
       480 {\lambda }^2 {c_1} {c_3} {{c_4 ^2}} \right.\cr
&\hspace{4.0cm}{}\left.{} + 
       768 {\pi }^2 \lambda  {{c_3 ^2}} {{c_4 ^2}} - 
       960 {\lambda }^2 {{c_4 ^3}} - 
       3072 {\pi }^2 \lambda  {c_2} {{c_4 ^3}} - 
       16384 {\pi }^4 {{c_4 ^4}} \right) \,,\cr
a_{-2}&=  \frac{3}
     {4096 {\pi }^4 {{c_4}}^{{5/2}}} \left( 5 {\lambda }^2 {{c_3 ^3}} - 
       20 {\lambda }^2 {c_2} {c_3} {c_4} + 
       40 {\lambda }^2 {c_1} {{c_4 ^2}} + 
       128 {\pi }^2 \lambda  {c_3} {{c_4 ^2}} \right) \,,\cr
a_{-3}&=  \frac{3 }{2048 {\pi }^4 {{c_4}}^{{3/2}}}\left( -5 {\lambda }^2 {{c_3 ^2}} + 20 {\lambda }^2 {c_2} {c_4} + 128 {\pi }^2 \lambda  {{c_4 ^2}} \right) \,,\cr
a_{-4}&=  \frac{15 {\lambda }^2 {c_3}}{512 {\pi }^4 {\sqrt{{c_4}}}}\,,\cr
 a_{-5}&= \frac{15 {\lambda }^2 {\sqrt{{c_4}}}}{256 {\pi }^4}\,,\nonumber 
\end{align}
and the normalization condition fixes $a_{1}$ as
\begin{equation}
a_{1}=\f{1}{2}-\alpha\,,
\end{equation}
where $\alpha$ is the filling fraction.  Putting all the {\sl ingredients} into the {\sl pot} (\ref{anomalous_dimension}), we obtain that 
\begin{align}
\frac{\gamma}{L} &=  \f{\tlambda}{8\pi^{2}}\left( \frac{{{c_3^2}}}{16{{c_4^2}}} - 
     \frac{{c_2}}{4{c_4}} + 
     \frac{{a_1}}{{\sqrt{{c_4}}}} \right)  \cr  
   &\cr
   &\hspace{-0.5cm}{}+ \f{\tlambda^{2}}{128\pi^{4}}\left( \frac{45 {{c_3^4}}}
      {256 {{c_4^4}}} - 
     \frac{27 {c_2} {{c_3^2}}}
      {32 {{c_4^3}}} + 
     \frac{3 {a_1} {{c_3^2}}}
      {8 {{c_4}}^{{5/2}}} + 
     \frac{9 {{c_2^2}}}{16 {{c_4^2}}} + 
     \frac{{c_1} {c_3}}{{{c_4^2}}} - 
     \frac{{a_1} {c_2}}
      {2 {{c_4}}^{{3/2}}} - 
     \frac{7}{4 {c_4}} \right) \cr
     &\cr
   &\hspace{-0.5cm}{}+ \f{\tlambda^{3}}{1024\pi^{6}}\left( \frac{285 {{c_3}}^6}
      {1024 {{c_4}}^6} - 
     \frac{467 {c_2} {{c_3^4}}}
      {256 {{c_4}}^5} + 
     \frac{35 {a_1} {{c_3^4}}}
      {128 {{c_4}}^{{9/2}}} + 
     \frac{195 {{c_2^2}} {{c_3^2}}}
      {64 {{c_4^4}}} + 
     \frac{2{c_1} {{c_3^3}}}{ {{c_4^4}}}- 
     \frac{15 {a_1} {c_2} {{c_3^2}}}
      {16 {{c_4}}^{{7/2}}} - 
     \frac{13 {{c_2^3}}}{16 {{c_4^3}}} \right.\cr
      &\cr
     &\hspace{0.0cm}\left.{}- 
     \frac{9 {c_1} {c_2} {c_3}}
      {2 {{c_4^3}}}  - 
     \frac{39 {{c_3^2}}}{16 {{c_4^3}}} + 
     \frac{3 {a_1} {{c_2^2}}}
      {8 {{c_4}}^{{5/2}}} + 
     \frac{3 {a_1} {c_1} {c_3}}
      {4 {{c_4}}^{{5/2}}} + 
     \frac{21 {{c_1^2}}}{16 {{c_4^2}}} + 
     \frac{3 {c_2}}{{{c_4^2}}} - 
     \frac{{a_1}}{2 {{c_4}}^{{3/2}}}
     \right)  {\tlambda }^3 +\O(\tlambda^{4})\,.
\end{align}

\subsubsection*{Symmetric Two-Cut Case}
Now we consider the symmetric two-cut cases, where the Riemann surface (\ref{Riemann_surface}) is given by
\begin{equation}\label{symmetric_Riemann_surface}
y^{2}=\ko{x^{2}-a^{2}}\ko{x^{2}-b^{2}}\,,
\end{equation}
with two cuts $(a,b)$ and $(-b,-a)$ on a double cover of the complex plane.  The differential (\ref{differential}) now reduces to
\begin{align}
d  p&=-\f{d  x}{\sqrt{\ko{b^{2}-x^{2}}\ko{x^{2}-a^{2}}}}\left[ \f{1}{2}-\alpha-\f{ab}{2x^{2}}
+\f{3\tlambda}{16\pi^{2}}\ko{-\f{ab}{x^{4}}+\f{1}{2x^{2}}\f{a^{2}+b^{2}}{ab}}\right.\cr
&\hspace{3.5cm}\left.{}+ \f{15\tlambda^{2}}{256\pi^{4}}\ko{-\f{ab}{x^{6}}+\f{1}{2x^{4}}\f{a^{2}+b^{2}}{ab}+\f{1}{8x^{2}}\f{\ko{a^{2}-b^{2}}^{2}}{a^{3}b^{3}} } \right]\,,
\end{align}
and the periods of $d  p$ can be expressed through the complete integrals of the first and the second kind.  We assign the mode number $n$ and $-n$ to the two cuts, normalizing the $\B$-period of $d  p$ to $4\pi n$.  We also put a condensate of density $m$ between the cuts, giving the $\A$-period $2\pi m$, i.e.,
\begin{equation}
\oint_{\A} d  p=2\pi m\,,\quad \qquad \oint_{\B} d  p=4\pi n\,.
\end{equation}
Converting them into the standard Legendre form with the help of the elliptic
integral formulae listed in Appendix \ref{app:Elliptic Integrals}, we obtain the following expressions: 
\begin{align}
2\pi i m&=\f{1}{b\sqrt{r}}\left[ \ko{1-2\alpha}\sqrt{r}\,\K{1-r}-\E{1-r}
\right]\cr
&\hspace{1.0cm}{}+\f{\tlambda}{16\pi^{2}}\f{1}{b^{3}r\sqrt{r}}\left[
2r\K{1-r} -\ko{1+r}\E{1-r}\right]\cr
&\hspace{1.0cm}{}+\f{3\tlambda^{2}}{1024\pi^{4}}\f{1}{b^{5}r^{2}\sqrt{r}}
\left[4r\ko{1+r}\K{1-r}
-\ko{3+2r+3r^{2}}\E{1-r}\right]\,,\label{A-cycles}\\ 2\pi
n&=\f{1}{b\sqrt{r}}\left[ \ko{1-2\alpha}\sqrt{r}\,\K{r}-\E{r}-\K{r}
\right]\cr
&\hspace{1.0cm}{}+\f{\tlambda}{16\pi^{2}}\f{1}{b^{3}r\sqrt{r}}\left[
\ko{1+r}\E{r} -\ko{1-r}\K{r}\right]\cr
&\hspace{1.0cm}{}+\f{3\tlambda^{2}}{1024\pi^{4}}\f{1}{b^{5}r^{2}\sqrt{r}}
\left[\ko{3+2r+3r^{2}}\E{r}
-\ko{3-2r-r^{2}}\K{r}\right]\,, \label{B-cycles}
\end{align}
where the moduli $r$ is defined by
\begin{equation}
r=\f{a^{2}}{b^{2}}\,. 
\end{equation}
Let us expand both the moduli $r$ and one of the endpoints $b$ 
in powers of $\tlambda$:
\begin{align}
r&=r_0+r_1\tlambda+r_2\tlambda^{2}+\O(\tlambda^3)\,,\\
b&=b_0+b_1\tlambda+b_2\tlambda^{2}+\O(\tlambda^3)\,,
\end{align}
and eliminate the unwanted parameters by using 
Eqs.\,(\ref{A-cycles}) and (\ref{B-cycles}).  Then from the $\O(\tlambda)$ relation
\begin{equation}
\label{fraction}
1-2\alpha=\f{1}{\sqrt{r_0}}\,\f{n\E{1-r_0}
+i m\left[ \E{r_0}-\K{r_0} \right]}{n\K{1-r_0}-i  m\K{r_0}}\,,
\end{equation}
we can determine the moduli $r_0$, with which we can express $\gamma/L$ as
\begin{align}\label{formula}
\f{\gamma}{L}&=-\f{\tlambda}{2\pi^2}( n\widetilde{\rm \bf K}  -i  m {\rm \bf K})\big[ -2n \widetilde{\rm \bf E} -2i  m {\rm \bf E} +n\ko{1+r_0} \widetilde{\rm \bf K} +i  m \ko{1-r_0}{\rm \bf K}\big]\cr
&\quad {}-\f{\tlambda^2}{8\pi^4}(n \widetilde{\rm \bf K} -i  m {\rm \bf K})^3\big[ -4n\ko{1+r_0}\widetilde{\rm \bf E}-4i  m(1+r_0){\rm \bf E}\cr
&\quad \hspace{3.8cm}{}+n(1+6r_0+r_0^2)\widetilde{\rm \bf K}+i  m(3-2r_0-r_0^2){\rm \bf K}\big]\cr
&\quad {}-\frac{{\tlambda }^3}{4 {\pi }^6 }{( n {\widetilde {\rm \bf K}} - i    m {\rm \bf K}) }^5\Big\{ -
	 {\big[ n {\widetilde {\rm \bf E}} + i    m ( {\rm \bf E} - {\rm \bf K} )  \big] }^2 
			( 3 n {\widetilde {\rm \bf E}} +  3 i    m {\rm \bf E} -   n {\widetilde {\rm \bf K}} -  2 i   m {\rm \bf K} )\cr
&\quad \hspace{.6cm}{} 	  -\big[ n {\widetilde {\rm \bf E}} +  i    m ( {\rm \bf E} - {\rm \bf K} )  \big] \big[ 2 n^2 {{\widetilde {\rm \bf E}}}^2 - 2 m^2 {{\rm \bf E}}^2 + 3 n^2 {{\widetilde {\rm \bf K}}}^2 +  5 i    m n {\widetilde {\rm \bf K}} {\rm \bf K} + 6 m^2 {{\rm \bf K}}^2 \cr
&\quad \hspace{4.5cm}{}- m {\rm \bf E} 
		( 11 i     n  {\widetilde {\rm \bf K}} + 7 m {\rm \bf K} ) +  i    n {\widetilde {\rm \bf E}} ( 4 m {\rm \bf E} + 11 i     n  {\widetilde {\rm \bf K}} + 7 m {\rm \bf K} )  
	\big] {{\rho }_0}  \cr
&\quad \hspace{.3cm}{}+ \big[ -3 n^3 {{\widetilde {\rm \bf E}}}^3 +  3 i     m^3  {{\rm \bf E}}^3 +  4 n^3 {{\widetilde {\rm \bf K}}}^3 + 6 i  m n^2  {{\widetilde {\rm \bf K}}}^2  {\rm \bf K}+ 13 m^2 n {\widetilde {\rm \bf K}}  {{\rm \bf K}}^2 -  6 i    m^3 {{\rm \bf K}}^3\cr
&\quad \hspace{1.0cm}{}+   n^2 {{\widetilde {\rm \bf E}}}^2 	(  -9 i    m {\rm \bf E} +   11 n {\widetilde {\rm \bf K}} - 2 i     m {\rm \bf K})\cr
&\quad \hspace{1.0cm}{}		+ m^2 {{\rm \bf E}}^2 		( -11 n {\widetilde {\rm \bf K}} + 2 i      m {\rm \bf K})  + i    m {\rm \bf E} 		( -18 n^2 {{\widetilde {\rm \bf K}}}^2 + 14 i     m n {\widetilde {\rm \bf K}}  {\rm \bf K} + 5 m^2 {{\rm \bf K}}^2 ) \cr
&\quad \hspace{1.0cm}{}+ n {\widetilde {\rm \bf E}} 			( 9 m^2 {{\rm \bf E}}^2 - 18 n^2 {{\widetilde {\rm \bf K}}}^2 +  14 i     m n {\widetilde {\rm \bf K}} {\rm \bf K} + 5 m^2 {{\rm \bf K}}^2 + 2 m {\rm \bf E} 
								(  11 i     n {\widetilde {\rm \bf K}} + 2 m {\rm \bf K} )  
							)
        \big]  {{{\rho }_0}}^2\cr
&\quad \hspace{.3cm}{}+ 
      	( n {\widetilde {\rm \bf K}} -  i    m {\rm \bf K} )  
       	\big[ n^2 {{\widetilde {\rm \bf E}}}^2 - m^2 {{\rm \bf E}}^2 + 4 n^2 {{\widetilde {\rm \bf K}}}^2  - 5 i   m n {\widetilde {\rm \bf K}} {\rm \bf K} - 2 m^2 {{\rm \bf K}}^2  \cr
&\quad \hspace{3.0cm}{} - m {\rm \bf E} 		( 3 i     n {\widetilde {\rm \bf K}} +  m {\rm \bf K} )+ i    n {\widetilde {\rm \bf E}} 	( 2 m {\rm \bf E} + 3 i   n {\widetilde {\rm \bf K}} + m {\rm \bf K}) 
	\big]  {{{\rho }_0}}^3 
\Big\}\cr
&\quad \hspace{.3cm}{}\times \Big\{(n {\widetilde {\rm \bf E}} + i    m {\rm \bf E} - n {\widetilde {\rm \bf K}} )  
    [ n {\widetilde {\rm \bf E}} + i    m ( {\rm \bf E} - {\rm \bf K} )  + 
   ( - n {\widetilde {\rm \bf K}}   +  i    m {\rm \bf K} )  {{\rho }_0}]\Big\}^{-1}
+\O(\tlambda^4)\,.
\end{align}
Here we have denoted $\K{r_0}$, $\E{r_0}$, $\K{1-r_0}$ and $\E{1-r_0}$ as
${\rm \bf K}$, ${\rm \bf E}$, $\widetilde{\rm \bf K}$ and $\widetilde{\rm \bf E}$,
respectively, for simplicity.  Equation (\ref{formula}) is the general elliptic,
three-loop formula with general
periods $(n,m)$.   It can be used to compute the anomalous dimension 
in the planar $SU(2)$ sector.
It is easy to check that the formula (\ref{formula}) correctly
reproduces the known results of the double contour and the
imaginary root solution.  For example, when we set $\ko{n,m}=\ko{1,0}$, 
it reduces to the double contour (or `folded') case.  It exactly
agrees with Eq.\,(51) in \cite{SS} at the three-loop level.  As
another case, $\ko{n,m}=\ko{0,2}$ corresponds to the imaginary root
(or `circular') case, reproducing Eq.\,(69) in \cite{SS}.  Thus we
have rederived the known results for the elliptic
sector. \\

Finally we will make some comments on the relation between the
double contour and the imaginary root solution.  It can be
shown that the two allowed solutions are considered as different
branches of the same function.  To see this, let us set
$m=-n$, then the formula (\ref{formula}) reduces to
\begin{align}
\label{formula2}
\f{\gamma}{L}&=-\f{n^{2}\tlambda}{2\pi^2}\ko{
\widetilde{\rm \bf K} +i {\rm \bf K}}\ko{ -2 \widetilde{\rm \bf E} +2i {\rm \bf E}
+\ko{1+r_0} \widetilde{\rm \bf K} -i \ko{1-r_0}{\rm \bf K}}\cr
&{}-{\f{n^{4}\tlambda^2}{8\pi^4}( \widetilde{\rm \bf K} +i {\rm \bf K})^3[
-4\ko{1+r_0}\widetilde{\rm \bf E}+4i (1+r_0){\rm
E}+(1+6r_0+r_0^2)\widetilde{\rm \bf K}-i (3-2r_0-r_0^2){\rm \bf K}]}\cr 
&{}-{\frac{n^{6}{\tlambda }^3}{4 {\pi }^6 }{( {\widetilde {\rm \bf K}} + i
{\rm \bf K}) }^5\Big\{ - {[ {\widetilde {\rm \bf E}} - i ( {\rm \bf E} - {\rm \bf K} ) ]
}^2 ( 3 {\widetilde {\rm \bf E}} - 3 i {\rm \bf E} - {\widetilde {\rm \bf K}} + 2 i
{\rm \bf K} ) -[ {\widetilde {\rm \bf E}} - i ( {\rm \bf E} - {\rm \bf K} ) ] } \cr
&\hspace{.6cm}{ \times [ 2 {{\widetilde {\rm \bf E}}}^2 - 2 {{\rm
E}}^2 + 3 {{\widetilde {\rm \bf K}}}^2 - 5 i {\widetilde {\rm \bf K}} {\rm \bf K} +
6 {{\rm \bf K}}^2 + {\rm \bf E} ( 11 i {\widetilde {\rm \bf K}} - 7 {\rm \bf K} ) + i
{\widetilde {\rm \bf E}} ( -4 {\rm \bf E} + 11 i {\widetilde {\rm \bf K}} - 7 {\rm
K} ) ] {{\rho }_0}} \cr
&\hspace{.3cm}{{}+ [ -3  {{\widetilde {\rm \bf E}}}^3 -  3 i       {{\rm \bf E}}^3 +  4  {{\widetilde {\rm \bf K}}}^3 - 6 i    {{\widetilde {\rm \bf K}}}^2  {\rm \bf K}+ 13  {\widetilde {\rm \bf K}}  {{\rm \bf K}}^2 +  6 i     {{\rm \bf K}}^3+    {{\widetilde {\rm \bf E}}}^2 	(  9 i     {\rm \bf E} +   11  {\widetilde {\rm \bf K}} + 2 i      {\rm \bf K}) } \cr
&\hspace{.6cm}{{}		+  {{\rm \bf E}}^2 		( -11  {\widetilde {\rm \bf K}} - 2 i       {\rm \bf K})  - i     {\rm \bf E} 		( -18  {{\widetilde {\rm \bf K}}}^2 - 14 i      {\widetilde {\rm \bf K}}  {\rm \bf K} + 5  {{\rm \bf K}}^2 )  }\cr
&\hspace{.6cm}{{}+  {\widetilde {\rm \bf E}} 			( 9  {{\rm \bf E}}^2 - 18  {{\widetilde {\rm \bf K}}}^2 -  14 i      {\widetilde {\rm \bf K}} {\rm \bf K} + 5  {{\rm \bf K}}^2 - 2  {\rm \bf E} 
								(  11 i      {\widetilde {\rm \bf K}} - 2  {\rm \bf K} )  
							)
        ]  {{{\rho }_0}}^2} \cr
&\hspace{.3cm}{{}+ 
      	(  {\widetilde {\rm \bf K}} + i     {\rm \bf K} )  
       	\left[  {{\widetilde {\rm \bf E}}}^2 -  {{\rm \bf E}}^2 + 4  {{\widetilde {\rm \bf K}}}^2  + 5 i    {\widetilde {\rm \bf K}} {\rm \bf K} - 2  {{\rm \bf K}}^2  
	       + {\rm \bf E} 		( 3 i      {\widetilde {\rm \bf K}} - {\rm \bf K} )+ i     {\widetilde {\rm \bf E}} 	( -2  {\rm \bf E} + 3 i    {\widetilde {\rm \bf K}} - {\rm \bf K}) 
	\right]  {{{\rho }_0}}^3 
\Big\}}\cr
&\hspace{.3cm}{\times \Big\{(n {\widetilde {\rm \bf E}} - i     {\rm \bf E} -  {\widetilde {\rm \bf K}} )  
    [  {\widetilde {\rm \bf E}} - i     ( {\rm \bf E} - {\rm \bf K} )  + 
   ( -  {\widetilde {\rm \bf K}}   -  i     {\rm \bf K} )  {{\rho }_0}]\Big\}^{-1}}
+\O(\tlambda^4)\,.
\end{align}
Complete elliptic integrals $\E{x}$ and $\K{x}$ have a logarithmic
branch-point at $x=1$, which turns out to be $r_{0}=0$ and $r_{0}=1$ in
our case. Taking account of the discontinuities,
which come from the analytical continuation through the singular points, 
one can show that the double contour solution is obtained as a branch
continued past $r_{0}=0$, while the imaginary root solution as a
branch continued past $r_{0}=1$. The former corresponds to a folded
rotating string folding $n$ times onto itself, and the latter to a
circular rotating string winding $2n$ times around itself.

\section{Elliptic Integrals and Elliptic Functions\label{app:Elliptic Integrals}}

	\subsection{Complete Elliptic Integrals}

\paragraph{Definition}
Our convention for the complete elliptic integrals of the first and the second kind are as follows:
\begin{alignat}{3}
\K{r} &\equiv \int_{0}^{1} \frac{d x}{\sqrt{\ko{1-x^2}\ko{1-rx^2}}}&{}&=\int_{0}^{\pi/2} \frac{d \varphi}{\sqrt{1-r\sin^{2}\varphi}}\,,\\
\E{r} &\equiv \int_{0}^{1} d x\,\sqrt{\frac{1-rx^2}{1-x^2}}&{}&=\int_{0}^{\pi/2} d \varphi\,\sqrt{1-r\sin^{2}\varphi}\,.
\end{alignat}
These are related to the hypergeometric functions as
\begin{equation}
 {}_2{\mathrm F}_1\ko{\mbox{$\f{1}{2},\f{1}{2};1;x$}}=\f{2}{\pi}\,\K{x}\, ,\qquad 
 {}_2{\mathrm F}_1\ko{\mbox{$-\f{1}{2},\f{1}{2};1;x$}}=\f{2}{\pi}\,\E{x}\, .
\end{equation}

\paragraph{Modular Transformation}
The complete elliptic integrals transform under $r\to 1/r$ as follows:
\begin{alignat}{3}
&\K{\f{1}{r}}&{}&=\sqrt{r}\left(\K{r}-i \K{1-r}\right)\,, \label{mod.trans-1}\\
&\E{\f{1}{r}}&{}&={\sqrt{r}}\left[ {\E{r}+i \E{1-r}-\ko{1-r}\K{r}-i r\K{1-r}}\right]\,,\label{mod.trans-2}\\
&\K{1-\f{1}{r}}&{}&=\sqrt{r}\,\K{1-r}\,,\label{mod.trans-3}\\
&\E{1-\f{1}{r}}&{}&=\mbox{$\f{1}{\sqrt{r}}$}\,{\E{1-r}}\,.\label{mod.trans-4}
\end{alignat}

\paragraph{Legendre Relation}
\begin{equation}
\K{r}\E{1-r}-\K{r}\K{1-r}+\E{r}\K{1-r}=\f{\pi}{2}\,.\label{Legendre Relation}
\end{equation}

	\subsection{Jacobi Elliptic Function\label{elliptic function}}

\paragraph{Definition}
The elliptic function $\mathrm{sn}\ko{u,r}$ is defined by
\begin{equation}
u\equiv \int_{0}^{\mathrm{sn}\ko{u,r}}\frac{d x}{\sqrt{\ko{1-x^2}\ko{1-rx^2}}}\,.
\end{equation}
The other two elliptic functions $\mathrm{cn}\ko{u,r}$ and $\mathrm{dn}\ko{u,r}$ are related to $\mathrm{sn}\ko{u,r}$ as 
\begin{align}
1&=\mathrm{cn}^{2}\ko{u,r}+\mathrm{sn}^{2}\ko{u,r}\,,\\
1&=\mathrm{dn}^{2}\ko{u,r}+r\,\mathrm{sn}^{2}\ko{u,r}\,,
\end{align}
and they are defined by
\begin{align}
\mathrm{cn}\ko{u,r}&=\sqrt{1-\mathrm{sn}^{2}\ko{u,r}}\,,\\
\mathrm{dn}\ko{u,r}&=\sqrt{1-r\,\mathrm{sn}^{2}\ko{u,r}}\,.
\end{align}
The sign of the square root is determined so that they are positive definite in the interval $u\in \ko{0,\K{r}}$.

\paragraph{Periodicity}
There are following relation between the complete elliptic integral of the first kind $\K{r}$ and the Jacobi elliptic functions $\mathrm{sn}\ko{u,r}$, $\mathrm{cn}\ko{u,r}$:
\begin{alignat}{5}
\mathrm{sn}\ko{u+2\,\K{r},r}&=-\mathrm{sn}\ko{u,r}\,,\label{periodicity-sn}\\
\mathrm{cn}\ko{u+2\,\K{r},r}&=-\mathrm{cn}\ko{u,r}\,,\label{periodicity-cn}\\
\mathrm{dn}\ko{u+2\,\K{r},r}&=\mathrm{dn}\ko{u,r}\,,\label{periodicity-dn}
\end{alignat}
i.e., $4\,\K{r}$ is the period of $\mathrm{sn}\ko{u,r}$ and $\mathrm{cn}\ko{u,r}$, while the period of $\mathrm{dn}\ko{u,r}$ is $2\,\K{r}$.

	\subsection{Some Useful Integral Formulae\label{app:integral formula}}

Elliptic integral formulae listed below are useful in the intermediate calculation in Appendix \ref{app:3-loop formula}.  The moduli parameter $r$ is defined as $r=a^{2}/b^{2}$.  
Then the following integral formulae are useful for computing the ${\A}$-cycle of the $(n,m)$-elliptic solution:
\begin{eqnarray}
&&\int_{a}^{b}\f{d  x}{\sqrt{\ko{b^{2}-x^{2}}\ko{x^{2}-a^{2}}}}=\f{1}{b}\,\K{1-r}\,,\\
&&\int_{a}^{b}\f{d  x}{x^{2}\sqrt{\ko{b^{2}-x^{2}}\ko{x^{2}-a^{2}}}}=\f{1}{a^{2}b}\,\E{1-r}\,,\\
&&\int_{a}^{b}\f{d  x}{x^{4}\sqrt{\ko{b^{2}-x^{2}}\ko{x^{2}-a^{2}}}}=\f{1}{3a^{4}b^{3}}\left[2\ko{a^{2}+b^{2}}\E{1-r}-a^{2}\K{1-r}\right]\,,\\
&&\int_{a}^{b}\f{d  x}{x^{6}\sqrt{\ko{b^{2}-x^{2}}\ko{x^{2}-a^{2}}}}=\f{1}{15a^{6}b^{5}}\left[\ko{8a^{4}+7a^{2}b^{2}+8b^{4}}\E{1-r}\right.\cr
&&\hspace{8.0cm}\left.{}-4a^{2}\ko{a^{2}+b^{2}}\K{1-r}\right]\,,
\end{eqnarray}
and the same for the ${\B}$-cycle:
\begin{eqnarray}
&&\int_{b}^{\infty}\f{d  x}{\sqrt{\ko{x^{2}-a^{2}}\ko{x^{2}-b^{2}}}}=\f{1}{b}\,\K{r}\,,\\
&&\int_{b}^{\infty}\f{d  x}{x^{2}\sqrt{\ko{x^{2}-a^{2}}\ko{x^{2}-b^{2}}}}=\f{1}{a^{2}b}\left[\K{r}-\E{r}\right]\,,\\
&&\int_{b}^{\infty}\f{d  x}{x^{4}\sqrt{\ko{x^{2}-a^{2}}\ko{x^{2}-b^{2}}}}=\f{1}{3a^{4}b^{3}}\,\left[\ko{a^{2}+2b^{2}}\K{r}-2\ko{a^{2}+b^{2}}\E{r}\right]\,,\\
&&\int_{b}^{\infty}\f{d  x}{x^{6}\sqrt{\ko{x^{2}-a^{2}}\ko{x^{2}-b^{2}}}}=\f{1}{15a^{6}b^{5}}\,\left[\ko{4a^{4}+3a^{2}b^{2}+8b^{4}}\K{r}\right.\cr
&&\hspace{8.0cm}\left.{}-\ko{8a^{4}+7a^{2}b^{2}+8b^{4}}\E{r}\right]\,.
\end{eqnarray}

%


\end{document}